\documentclass[sigconf,balance=false]{acmart}
\AtBeginDocument{%
  }

\copyrightyear{2026}
\acmYear{2026}
\setcopyright{cc}
\setcctype{by}
\acmConference[SoCC '26]{ACM Symposium on Cloud Computing V.1}{November 18--20, 2026}{Singapore, Singapore}
\acmBooktitle{ACM Symposium on Cloud Computing V.1 (SoCC '26), November 18--20, 2026, Singapore, Singapore}
\acmDOI{10.1145/3815789.3827950}
\acmISBN{979-8-4007-2735-1/2026/11}

\usepackage{amsmath}

\usepackage[capitalize, noabbrev, nameinlink]{cleveref}
\newcommand{\beginbsec}[1]{\noindent\textbf{#1. \hspace{2pt}}}

\usepackage{tikz}
\usepackage{pgfplots}
\pgfplotsset{compat=1.18}
\usepackage{pgfplotstable}
\usepgfplotslibrary{groupplots}
\usepackage{stfloats}

\usepackage{algorithm}
\usepackage{algpseudocode}
\usepackage[normalem]{ulem} 
\usepackage{soul}
\usepackage{pifont}
\usepackage{tabularx}
\usepackage{multirow} 
\definecolor{OliveGreen}{rgb}{0,0.6,0}
\usepackage{arydshln}
\usepackage{pgfplots}
\usepackage{pgfplotstable}
\usetikzlibrary{positioning,shapes.geometric, positioning, arrows.meta}
\usetikzlibrary{patterns}
\usepackage{subcaption}
\usepackage{booktabs} 
\pgfplotsset{compat=1.18}
\usepgfplotslibrary{groupplots}

\newcommand{\squishlistnum}{  
 \newcounter{qcounter}

 \begin{list}{\roman{qcounter})~}{\usecounter{qcounter}}
  { \setlength{\itemsep}{0pt}
     \setlength{\parsep}{0pt}
     \setlength{\topsep}{0pt}
     \setlength{\partopsep}{0pt}
     \setlength{\leftmargin}{0em}
     \setlength{\labelwidth}{0em}
     \setlength{\labelsep}{0em} } }

\newcommand{\squishlist}{ 
 \begin{list}{$\bullet$}
  { \setlength{\itemsep}{3pt}
     \setlength{\parsep}{0pt}
     \setlength{\topsep}{3pt}
     \setlength{\partopsep}{0pt}
     \setlength{\leftmargin}{1em}
     \setlength{\labelwidth}{1em}
     \setlength{\labelsep}{0.5em} } }

\newcommand{\squishend}{
  \end{list}  }

\newcommand{\coolName}{Spandana}
\newcommand{\slb}{SLO Director}
\newcommand*\circled[1]{\tikz[baseline=(char.base)]{
            \node[shape=circle,draw,inner sep=1pt] (char) {#1};}}

\begin{document}


\title[Reconciling Strict SLOs with Low Cost under Fine-Grained Load Fluctuations]{Reconciling Strict SLOs with Low Cost \\ under Fine-Grained Load Fluctuations}

\author{Dilina Dehigama}
\affiliation{%
  \institution{University of Edinburgh}
  \country{}
  }

\author{Shyam Jesalpura}
\affiliation{%
  \institution{University of Edinburgh}
  \country{}
  }

\author{Zeyu Xu}
\affiliation{%
  \institution{University of Edinburgh}
  \country{}
  }

\author{Marton Nemeth}
\affiliation{%
  \institution{University of Edinburgh}
  \country{}
  }

\author{Shengda Zhu}
\affiliation{%
  \institution{The University of Edinburgh}
  \country{}
  }

\author{Marios Kogias}
\affiliation{%
  \institution{Imperial College London}
  \country{}
  }

\author{Boris Grot}
\affiliation{%
  \institution{University of Edinburgh}
  \country{}
  }

\renewcommand{\shortauthors}{Dehigama et al.}

\begin{abstract}

Cloud-based online services must contend with a load that exhibits significant fluctuations at sub-second granularity. Despite such load volatility, services must meet strict Service Level Objectives (SLOs), which forces a tradeoff between over-provisioning resources in order to meet SLO and achieving high resource utilization and cost efficiency. None of the existing approaches, which include reactive and proactive autoscalers, serverless (FaaS), and hybrid clusters combining virtual machines (VMs) and serverless workers, are able to reconcile this tension under fine-grained load fluctuation.

We introduce \coolName, an architecture that resolves this tension by decoupling SLO enforcement from cost optimization through a two-level control plane. A light-weight controller colocated with each application VM (the local level) steers each arriving request to the VM if the expected service time falls within the SLO target; otherwise, the request is forwarded to a stock FaaS layer (e.g., AWS Lambda). Doing so achieves high VM utilization and cost-efficiency while ensuring strict SLO compliance through the use of elastic serverless instances when needed. At the global level, \coolName's resource allocator determines the most-efficient VM provisioning considering both VM and serverless costs as well as traffic volatility. 
Our evaluation shows that Spandana exhibits extremely strict SLO adherence, consistently achieves CPU utilization in the range of 76-86\%, and reduces costs by 5-44\% over three SOTA baselines. 
\end{abstract}

\ccsdesc[500]{Computer systems organization~Cloud computing}

\keywords{Autoscaling, Resource Provisioning, Cloud
Computing}



\maketitle

\begingroup
\endgroup

\newpage
\section{Introduction}
\label{sec:introduction}

User-facing services running in the cloud are subjected to a highly volatile request load, with significant fluctuations at a second and sub-second granularity~\cite{shan2016microburstdatacentersobservations,workload-spikes}. A recent study on load balancing at YouTube reports that server load appearing relatively smooth at a minute granularity may fluctuate by over an order of magnitude at a second granularity~\cite{youtube}. Our analysis of a Twitter trace shows that second-scale load fluctuations exceed the average load by as much as 50\%.

Such fine-grained load fluctuations make resource provisioning difficult. With cloud-based services typically deployed on VMs, system administrators are faced with conflicting requirements of keeping VM utilization low in order to meet SLO but also of minimizing the number of VMs in order to control costs. Reconciling these competing objectives is a real-world challenge. Since SLO objectives generally take priority for the sake of user experience, deployments are commonly overprovisioned relative to the average load so that fluctuations can be absorbed without violating SLO. As a result, industry reports point to CPU utilization in production clusters often well below 50\%\cite{resource-central-sosp17,coach-azure-asplos-25}.

To keep up with bursty loads, cloud deployments rely on autoscaling mechanisms that seek to adjust capacity in response to load fluctuations. Reactive autoscalers, such as the Kubernetes (k8s) Horizontal Pod Autoscaler (HPA)~\cite{kubernetes-hpa}, are designed to detect cases when utilization is above a preconfigured threshold for a sustained period of time, in which case they react by bringing more VM capacity online. Problematically, detection and bringing new VMs online can take minutes \cite{6253534}, which means such autoscalers cannot respond effectively to short-lived bursts. Proactive autoscalers aim to predict demand and scale resources in advance\cite{roy2011efficient, madu, ali2023hybrid, qiu2023aware, tong2024burstaware, e3former2025online}.
In practice, proactive systems operate at coarse time scales, such as minute-level intervals, to avoid instability.
As a result, fine-grained load fluctuations are either missed or require overprovisioning of resources leading to high deployment cost and low utilization. What both reactive and predictive schemes lack is fine-grained elasticity that can naturally accommodate fine-grained load fluctuations.

Function-as-a-Service (FaaS), or serverless computing, offers tremendous resource elasticity, making it attractive for volatile loads. Alas, as our studies, and that of others~\cite{libra, lambada}, show, serverless is several times more expensive than VMs for volume processing. A number of works have proposed augmenting VMs with serverless; the majority of these engage serverless instances only when faced with unexpected surges in load and/or to bridge the throughput gaps during new VMs being launched~\cite{pixel, cackle, splitserve, spock, mark, sponge, feat, hydra}. Such schemes do not help with fine-grained load fluctuations. Libra~\cite{libra} goes beyond these works by {\em continuously} offloading a portion of the traffic to serverless in order to balance SLO and cost. However, as our evaluation reveals, Libra's resource allocation strategy tries to provision for both SLO and cost in advance, resulting in a poor allocation under volatile traffic that compromises on cost in order to meet SLO.

Our work directly targets the tension between strict SLO, high resource utilization and low cost. We introduce \coolName, a fresh take on combining the cost-efficiency of conventional VMs with an elastic compute substrate (FaaS in this work, but other elastic resource pools are possible) to ensure strict latency SLO under highly volatile load.

The key insight exploited in \coolName's design is the decoupling of SLO enforcement from cost optimization. To that end, \coolName{} employs a two-level architecture. At the local level, each application instance (e.g., a VM) makes a local decision as to whether an arriving request can meet its SLO target based on the instance's current utilization level and queue length of pending requests. If the expected service time of the arriving packet exceeds the latency budget, the packet is immediately forwarded to the serverless plane.
Crucially, in \coolName{}, a request is processed by a serverless instance {\em only if} it cannot be accommodated by a VM without violating SLO, thereby ensuring high VM utilization and cost efficiency with strict SLO adherence. The decision-making logic concerning whether to process a request on a VM or redirect to serverless has minimal CPU and memory footprint and is colocated with each application instance, thereby scaling naturally with deployment size and with no impact on existing intra-VM load balancing~\cite{kubernetes-service-lb}.

At the global level, a deployment-wide resource optimizer periodically examines the load distribution across VMs and serverless instances and adjusts VM allocation so as to minimize overall deployment cost taking into account the extent of traffic fluctuations and serverless costs. Notably, the optimizer allocates {\em only} for overall cost without concern for SLO, which is enforced at \coolName's local level.

We evaluate \coolName{} against k8s HPA, a state-of-practice autoscaler, as well as AutoBurst~\cite{autoburst_socc} and Libra~\cite{libra}, state-of-the-art schemes for cost- and SLO-aware cluster provisioning that combine VMs with burstable (AutoBurst) and serverless (Libra) instances.
Our experiments show that \coolName{} comfortably meets strict latency SLO targets for all studied applications, and achieves desirable high CPU utilization of VM instances in the range of 76\%-86\% and reduces costs by 14-36\% compared to HPA, 5-27\% compared to AutoBurst and by 28-44\% compared to Libra whenever AutoBurst and Libra meet or only slightly exceed the SLO budget.

To summarize, we make the following contributions:

\squishlist
    \item Characterize how fine-grained load fluctuations present a unique challenge for VM-based application deployments, which must choose between cost-efficiency or SLO compliance.
    \item Existing resource allocation schemes force a choice between SLO adherence and cost-efficiency. Such schemes include both reactive and proactive autoscalers, serverless deployments and hybrid approaches combining VMs with serverless.
    \item \coolName{} decouples the concerns of SLO enforcement and cost efficiency into separate and dedicated control planes. SLO is enforced at each application instance by estimating whether an arriving packet can meet its latency SLO if it's processed by the instance; if not, it is redirected to FaaS, thus guaranteeing SLO adherence. Cost efficiency is achieved by a global controller that considers VM costs, serverless costs and traffic volatility.
    \item \coolName{} meets strict SLO targets while outperforming SOTA baselines on cost and CPU utilization.
\squishend

\section{Motivation}
\label{sec:motivation}

\subsection{The Bursty Nature of Traffic}
\label{sec:motivation:bursts}

The user-triggered load on the services in the cloud is known to exhibit significant fine-grained fluctuations, a key feature of which are so-called microbursts~\cite{shan2016microburstdatacentersobservations,workload-spikes}. Microbursts are characterized by sudden and short-lived spikes in load with a duration of tens of milliseconds or less. In the presence of microbursts, the load on individual services deployed in the cloud can fluctuate greatly at sub-second time scales, resulting in a volatile request pattern in which the average load at coarser time scales fails to represent the highly-variable actual load.

\begin{figure}[!t]
    \includegraphics[width=\linewidth]{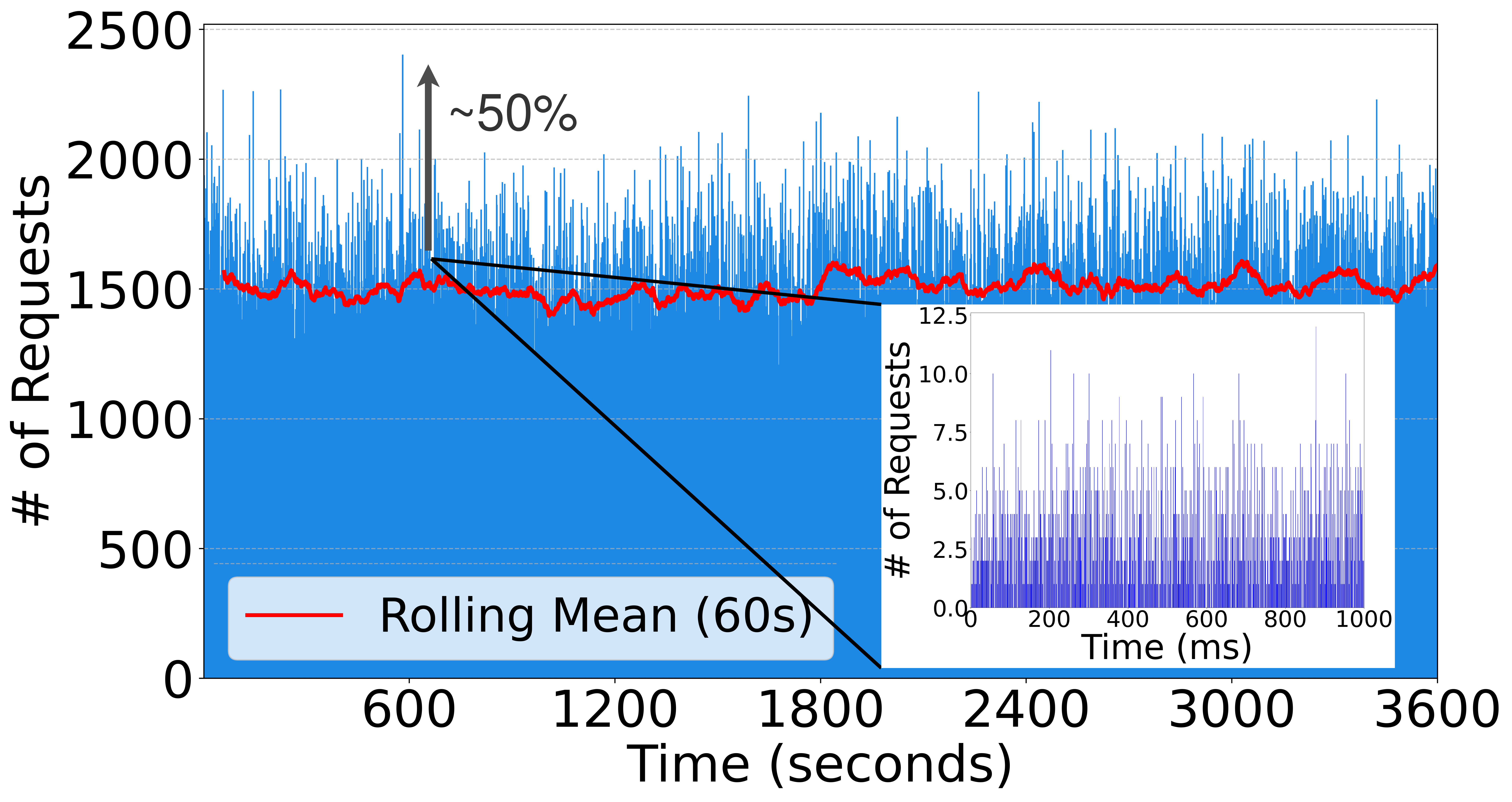}
    \caption{Production workload from Twitter showing fine-grained load fluctuations at 1-second granularity and zoomed-in figure showing fluctuations within a second}
    \label{fig:load}
\end{figure}

To illustrate this behavior, we analyze a one-hour segment of a production workload trace~\cite{twitter:trace} from Twitter (now, X), a popular user-facing cloud application. \cref{fig:load} shows the request rate at two granularities -- minute-scale (red) and second-scale (blue). At minute scale, computed as the rolling average over the previous 60 seconds, the load is quite stable, deviating from a trace-wide mean by at most 7.2\% across all intervals. In contrast, at second scale, the trace shows deviations from the mean of as much as 50\%.

Because the resolution of the data in the trace is limited to 1 second, we extrapolate the arrival rates within each 1-second interval following a Poisson process with an exponential distribution of inter-arrival times. The Poisson distribution is widely accepted as representative of human-generated traffic~\cite{serverless-wild,gallager2013stochastic}. \cref{fig:load} shows the resulting load with significant fluctuations within each second-long interval. Fluctuating pattern of load is not unique to this workload, with similar patterns having been observed in other large-scale systems, such as YouTube's production clusters \cite{youtube}.

\begin{figure*}[t]
    \centering
    \hfill
    \begin{subfigure}[b]{0.46\textwidth}
        \centering
        \includegraphics[width=\textwidth]{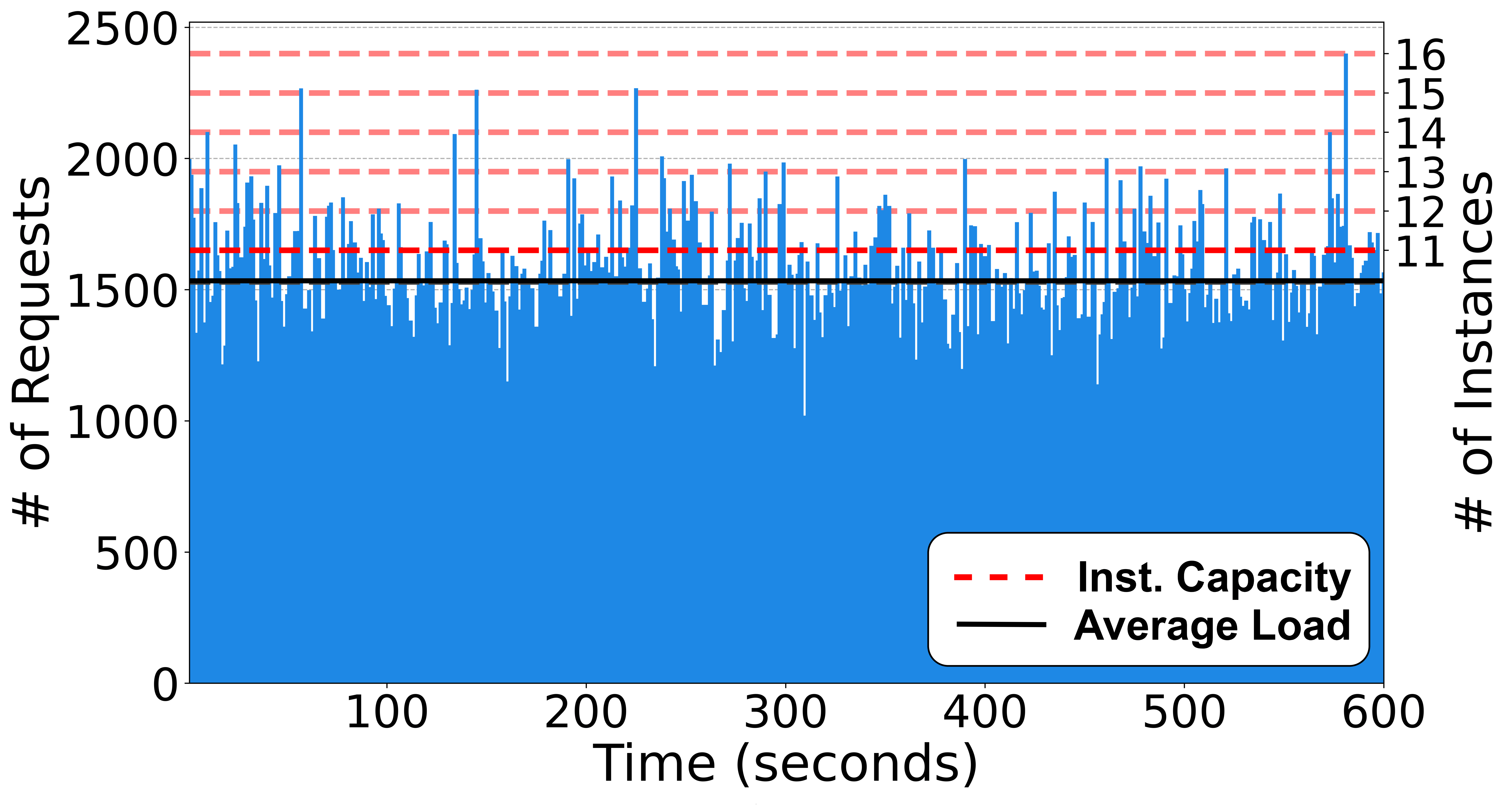}
   
        \caption{Combined request handling capacities vs cluster sizes.}
        \label{fig:load-n-instances}
    \end{subfigure}
    \hfill
    \begin{subfigure}[b]{0.43\textwidth} 
        \centering
        \begin{tikzpicture}
\begin{axis}[
    width=8cm,
    height=4.6cm,
    xlabel={SLO Violations (\%) (log scale)},
    ylabel={Average CPU Utilization (\%)},
    xmode=log,
    log ticks with fixed point,
    xmin=0.01, xmax=100,
    ymin=50, ymax=100,
    xtick={0.01, 0.1, 1, 10, 100},
    ytick={50,60,70,80,90,100},
    grid=major,
    legend style={
        at={(0.95,0.05)},
        anchor=south east,
        font=\footnotesize,
    },
    clip=false,
    extra x ticks={40},
    extra x tick labels={40},
    extra x tick style={
        tick style={red, thick},
        ticklabel style={red},
    },
]

\addplot[
    color=blue,
    mark=*,
    thick,
    nodes near coords,
    nodes near coords style={
        font=\small,
        anchor=north,
        yshift=2pt,
    },
    point meta=explicit symbolic
] coordinates {
    (38,90)   [11 inst.]
    (8.1,82.5) [12 inst.]
    (1.6,76)   [13 inst.]
    (0.9,69)   [14 inst.]
    (0.28,63)  [15 inst.]
    (0.02,60)  [16 inst.]
};

\addplot[
    color=red,
    dashed,
    thick,
] coordinates {(38,50) (38,100)};

\addplot[
    only marks,
    mark=triangle*,
    mark options={scale=1.2,fill=black},
] coordinates {(1,82)};
\node[anchor=east, xshift=-2pt,yshift=5pt, font=\small, align=right]
  at (axis cs:1,82) {11 inst.\\(excess removed)};

\end{axis}
\end{tikzpicture}
        \caption{CPU utilization vs SLO target violations (\%)}
        \label{fig:num-instances-cpu-slo}
    \end{subfigure}
    \hfill
    \caption{The provisioning trade-off for bursty workloads.}
    \label{fig:combined-provisioning-cpu}
\end{figure*}
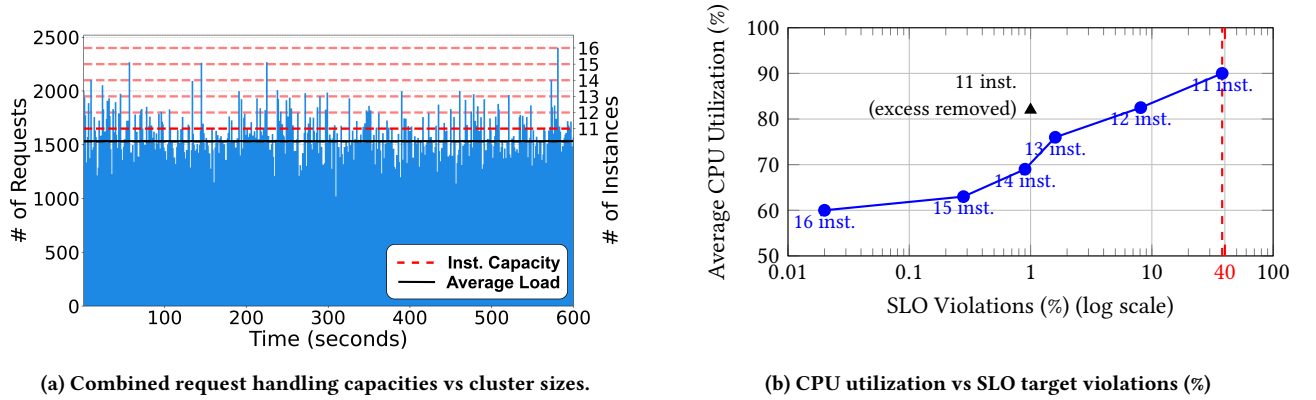

\subsection{Resource Provisioning for Fluctuating Load}
\label{sec:motivation:provisioning}

\begin{table}[h!]
\small
\centering
\begin{tabular}{lcccc}
\toprule
\textbf{System} & \textbf{SLO Vio. (\%)} & \textbf{Cost (¢)} & \textbf{Avg \# VM} & \textbf{Max \# VM}\\
\midrule
HPA-cost      & 3.5   & 9.2                & 12.1 & 15 \\
HPA-slo       & 0.29  & 10.0               & 14.0 & 14 \\
Serverless    & 0.02  & 30.9 & - & - \\
Libra         & 0.84  & 12.7 (8.9 / 3.8)  & 12.4 & 13 \\
\bottomrule
\end{tabular}
\caption{Comparison of SLO violations, cost, and number of VM instances across systems. For Libra, total cost is broken down as (VM / serverless).}

\label{fig:motivation:sota}
\end{table}
For cloud applications, fine-grained fluctuations create a significant challenge of resource provisioning\footnote{In today's clouds, resources are typically provisioned at the granularity of VMs. Within a VM, individual services are deployed as containers. 
}. The challenge lies in balancing two competing objectives: allocating enough resources to meet strict SLOs, while simultaneously keeping resource utilization high to avoid the cost of idle capacity.

To showcase the provisioning challenge, we conducted an experiment using a 10-minute portion of the same load from \cref{fig:load} applied to an online service, the \textit{ratings-service} from the BookInfo application \cite{bookinfo}. The \textit{ratings-service} was deployed on a multi-node AWS Elastic Kubernetes Service cluster (EKS) with m5a.large VMs with 2 vCPUs and 8 GB of memory. 
One vCPU was allocated to each container instance of the service hosted on the VMs.
From here on, we refer to each container instance of the service as an \textit{application instance}.
First, we determined the maximum throughput of a single application instance  by benchmarking it with a uniform request load to find the highest request rate it could handle without violating its latency SLO target. 
The latency SLO target was defined as 10 times the processing time of a single request on an idle application instance. 

\cref{fig:load-n-instances} shows the load trace (blue) along with the aggregate throughput of the application instances (dashed red lines). As the figure shows, the minimum cluster size needed to handle the average load (just over 1500 RPS) is 11 application instances.
However, the frequent bursts clearly exceed the capacity of an 11 application-instance cluster, which leads to severe SLO violations. 
To handle every peak in the trace, a total of 16 application instances are needed which is 45\% more than the 11 application instances required to sustain the average load.
\cref{fig:num-instances-cpu-slo} illustrates the direct consequence of provisioning more application instances on CPU utilization of the cluster.
Provisioning for the average load with 11 application instances drives CPU utilization to a desirable high of 90\%, but at the cost of an unacceptable 40\% SLO violation rate. 
Conversely, provisioning for the peak load with 16 application instances reduces SLO violations to near-zero but at the cost of CPU utilization dropping to 60\%, leaving a large fraction of CPU capacity idle. 
The experiment reveals the provisioning challenge for fluctuating load: clusters must either run at high utilization and violate SLOs, or they must be overprovisioned to meet SLOs at the cost of under-utilization. 

In fact, the under-utilization problem is even more severe in large-scale production environments. Recent industry reports show that average cluster CPU utilization often remains well below 50\%. For example, a recent report analyzing thousands of production clusters across major cloud providers found that clusters hosting production workloads reach only about 10\% average CPU utilization \cite{cast-ai-report}. Similarly, studies of Azure VM clusters show that majority of VMs have an average CPU utilization
below 50\% \cite{resource-central-sosp17,coach-azure-asplos-25}. 

\textbf{Takeaway:} With bursty workloads, clusters face a dilemma: provisioning for load peaks avoids SLO violations but requires over-provisioned, incurring cost overheads and low utilization; provisioning for the average load leads to frequent SLO violations.

\section{State-of-the-Art Approaches Fall Short}
\label{sec:sota}

Current cloud provisioning strategies struggle to reconcile the conflicting goals of strict SLO compliance, high resource utilization, and low operational cost when facing fine-grained load fluctuations. We categorize the existing solutions into three predominant classes: autoscaling with VMs, pure serverless deployments, and a hybrid approach that combines VMs and serverless~\cite{libra}. Each category addresses part of the problem, but none handles fine-grained bursts without either sacrificing utilization or paying a high premium.

\subsection{Limitations of VM Autoscaling}
\label{sec:sota:autoscale}
Standard cloud deployments rely on autoscaling to adjust the number of deployed VMs in response to demand. 
Reactive autoscalers, such as the k8s HPA, monitor resource usage over a window of time and add capacity when utilization exceeds a threshold~\cite{kubernetes-hpa}. A number of works have proposed more advanced autoscaling strategies~\cite{padala2009automated, gandhi2012autoscale} but all are ultimately hamstrung by two limitations: (1) new VMs are slow to start, often requiring a minute or more~\cite{6253534}; and (2) the slow VM start times result in long decision intervals (10s of seconds) to avoid spinning up VMs for transient bursts.

Proactive autoscalers attempt to forecast demand in order to scale ahead of time, but such predictions typically operate at coarse, minute-level granularities to maintain stability~\cite{madu,Autopilot}. 
As a result, under highly fluctuating load at sub-second time scales, both reactive and proactive autoscalers must be tuned {\em either} for high SLO compliance, resulting in over-provisioning, low utilization and high cost, {\em or} for high utilization and low cost, leading to SLO violations due to microbursts.

We use HPA as a representative state-of-practice autoscaler and evaluate its effectiveness on the Ratings service from Bookinfo application~\cite{bookinfo} under the fluctuating load pattern from \cref{fig:load}.
The SLO is defined as a 99th-percentile tail latency target of 140\,ms, set at 10$\times$ the median latency of the service under light load. See \cref{sec:methodology} for details of the methodology.
\cref{fig:motivation:sota} illustrates the cost and performance of these approaches. The following subsections discuss the other techniques shown in the table. 

Tuning HPA for cost (HPA-cost) results in a significant rate of SLO violations of 3.5\% while utilizing an average of 12.1 VMs and peaking at 15 VMs. Conversely, tuning HPA for SLO (HPA-slo) decreases the SLO violation rate to 0.29\% but increases the average VM count to 14.0. The extra provisioning required to absorb fluctuations, increases the total cost to 10.03c compared to 9.18c for HPA-cost. Autoscaling approaches can easily reconcile resource utilization and SLO under load that varies only at coarser time intervals, such as minute-scale. However, when load fluctuates at sub-second granularity, which is the case in existing cloud services~\cite{youtube}, autoscalers must choose between SLO and resource utilization, the latter directly correlated with cost.
Lastly, we note that while our study focused on the state-of-practice HPA, other autoscalers face the same fundamental limitation due to the lack of fine-grained elasticity that prevents effective mitigation of sub-second load variance.
\subsection{Serverless: SLO at high cost}
\label{sec:sota:serverless}
Serverless computing, specifically Function-as-a-Service (FaaS) platforms like AWS Lambda, offers the fine-grained elasticity that VMs lack. 
Serverless\footnote{
  We use FaaS and serverless interchangeably. That is, any usage of serverless refers strictly to FaaS.}
provides extremely fast (100s of ms) startup, on-demand autoscaling, and a pay-for-use billing model -- features that are well-aligned with demands imposed by bursty loads. 

To assess the viability of a pure serverless approach, we execute the fluctuating workload on stock AWS Lambda functions. We do not use any performance optimizations, such as pre-warming the functions or keeping them alive through periodic pings; as such, cold start latencies and other inherent performance variabilities of FaaS are reflected in our measurements. 

As seen in \cref{fig:motivation:sota}, serverless comfortably meets the latency target, with 99.98\% of requests completing within the SLO threshold (recall that the target is 99\%). Such performance, however, comes at an exorbitant cost, exceeding SLO-compliant VM-based deployments by 3.1x.  
Thus, while serverless can address the SLO challenge, it fails the cost-efficiency requirement for high-volume services.

\subsection{Hybrid Compute to the Rescue?}
\label{sec:sota:hybrid}

As demonstrated above, serverless can provide high resource elasticity that is a good fit for fluctuating service loads and SLO targets. Meanwhile, VMs offer much better cost efficiency for volume processing. Not surprisingly, prior work has explored approaches that combine the two types of compute. 

A number of papers advocate for using serverless as a fallback resource during VM scale-out and/or when the load surges~\cite{pixel, cackle, splitserve, spock, mark, sponge, feat, hydra, microless}. The general theme is to serve the load on VMs, and engage serverless only when the load shifts and VMs cannot cope. 
Such schemes are effective for coarser-grained variations in load, but the high latency used to detect a load shift and the mechanisms used to redirect the traffic are not well-suited for fine-grained load fluctuations.

A more attractive approach for handling high load variability at sub-second time scales is to {\em continuously} use hybrid compute types to achieve both SLO compliance and cost-efficiency. The state-of-the-art in doing so is Libra~\cite{libra}. In each time interval, Libra processes a pre-determined fraction of the traffic on VMs and sends the rest to serverless.

We identify three critical limitations in the Libra design that lead to suboptimal performance.
First, Libra uses a fixed ratio the determines the fraction of traffic served by serverless. The user sets the parameter once with no mechanism to determine whether the chosen ratio is optimal at the current point in time.
Second, when steering the requests to VMs or serverless, Libra follows a policy based on aggregated request volume that is unaware of the instantaneous load on the VMs. Thus, when a microburst arrives, Libra may overload the VMs, resulting in SLO violations. To account for this, Libra overprovisions VMs, failing to capitalize on the elasticity offered by serverless.
Lastly, Libra's approach for determining the optimal number of VM instances from a cost/performance perspective only considers the total load in a previous time interval. As we show in the next section, such an approach is myopic and leads to a suboptimal resource allocation. Instead, the resource allocator must take load fluctuations into account since they affect SLO and the deployment must meet the SLO by having sufficient resources.

\cref{fig:motivation:sota} highlights the inefficiency of the allocation strategy employed by Libra. The system provisions an average of 12.4 VM instances compared to 14.0 for the HPA-slo configuration. The reduction of 1.6 instances fails to offset the cost overhead of serverless, resulting in a total cost 1.3x higher than the HPA-slo configuration.
With its rigid request distribution, limited awareness of microbursts, and myopic resource allocation, Libra fails to achieve both cost-efficiency and SLO compliance at once.

\section{Reconciling Strict SLO with Low Cost}
\label{sec:bridge}

The appeal of combining VMs with serverless to effectively handle volatile load lies in the former's cost-efficiency and the latter's resource elasticity. However, as shown in the previous section, existing works have not been able to reconcile the objective of strict SLO with that of low cost. Send too much bursty traffic to VMs, and SLO suffers; offload excessively to FaaS, and costs spike. 

We observe that a system perfectly reconciling strict SLO with low cost under volatile load must abide by the following principles: \\
{\bf P1.} Any request that {\em can be} serviced on a VM without violating SLO {\em should be} serviced on a VM. This ensures cost-efficiency with SLO compliance. \\
{\bf P2.} A corollary of the above is that any request that cannot be serviced on a VM within the latency SLO (and only such requests!) must be sent to serverless. 
Doing so ensures strict SLO compliance while minimizing cost overheads of FaaS. 

A naive implementation of the above principles would simply overprovision the pool of VMs to fully absorb the fluctuations. As shown in \cref{sec:motivation:provisioning}, however, such a deployment would be plagued by low average VM utilization and cost inefficiency. Thus, we introduce the third, and final, principle: \\
{\bf P3.} The number of allocated VMs must be provisioned in a way that minimizes total cost, taking serverless costs into account. 

\vspace{0.02in}
Guided by the principles above, we unveil three insights that naturally lead to a system architecture capable of achieving strict SLO compliance and high cost-efficiency under load variability.

{\em Insight 1. Decision as to whether a request should be processed by a VM or a serverless instance must be per-request and on-demand.} Coarse-grained decisions, such as Libra's interval-based approach to commit a particular amount of load to VMs and not engage serverless until that point, inevitably lead to SLO violations and/or inflated costs under volatile load (\cref{sec:sota}). Instead, principles P1 and P2 dictate that each request should have an opportunity to execute on a VM (for cost-efficiency), and only execute on serverless when SLO compliance is at risk.

{\em Insight 2. Just because a VM is operating at its peak capacity at a certain point in time does not mean a request should be offloaded to serverless.} Latency SLO is typically defined as a multiple of the average-case request service time in the absence of queueing and resource contention. By design, this implies that a certain degree of queueing is acceptable from the latency SLO standpoint. Thus, the mere fact that a VM is fully occupied at a given instant of time does not mean that an arriving request cannot be completed within its SLO budget. As dictated by queueing theory, it is the combination of the service rate (of the VM) and the number of pending (queued) requests that together determine the expected service time of the request if it were to wait. Only if the expected time to completion, including the queueing time, exceeds the SLO target, should the request be offloaded to serverless. 

{\em Insight 3. Knowledge of average or total load is insufficient to find a cost-optimal VM provisioning.} For instance, Libra examines the number of requests received in a time interval, uses that to decide on a fraction to be processed by VMs, and from that, determines the number of VM instances needed. Problematically, under volatile load, such an approach results in a non-optimal number of allocated VMs ultimately leading to inflated cost. 
Instead, the extent of load variability must be considered for optimal VM provisioning. 

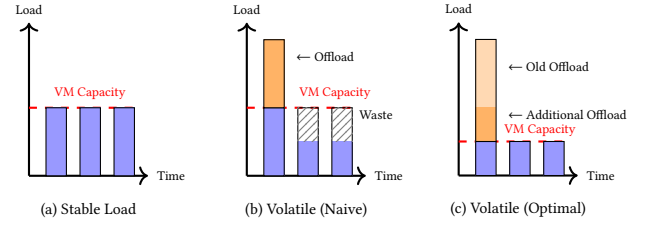
\begin{figure}[t!]
    \centering
    \begin{subfigure}[b]{0.32\linewidth}
        \centering
        \begin{tikzpicture}[scale=0.45]
            \draw[->, thick] (0,0) -- (3.5,0) node[right] {\tiny Time};
            \draw[->, thick] (0,0) -- (0,4.5) node[above] {\tiny Load};
            
            \draw[red, dashed, thick] (0, 2) -- (3.2, 2);
            \node[red, anchor=south west] at (0.5, 2) {\tiny VM Capacity};

            \foreach \x in {0.5, 1.5, 2.5} {
                \fill[blue!40] (\x,0) rectangle (\x+0.6, 2);
                \draw[black] (\x,0) rectangle (\x+0.6, 2);
            }
            \node[align=center, font=\scriptsize] at (1.75, -1) {(a) Stable Load};
        \end{tikzpicture}
    \end{subfigure}
    \hfill
    \begin{subfigure}[b]{0.32\linewidth}
        \centering
        \begin{tikzpicture}[scale=0.45]
            \draw[->, thick] (0,0) -- (3.5,0) node[right] {\tiny Time};
            \draw[->, thick] (0,0) -- (0,4.5) node[above] {\tiny Load};
            
            \draw[red, dashed, thick] (0, 2) -- (3.2, 2);
            \node[red, anchor=south west] at (1.3, 2) {\tiny VM Capacity};

            \fill[blue!40] (0.5,0) rectangle (1.1, 2);
            \draw[black] (0.5,0) rectangle (1.1, 2);
            \fill[orange!60] (0.5,2) rectangle (1.1, 4); 
            \draw[black] (0.5,2) rectangle (1.1, 4);
            \node[font=\tiny] at (2.3, 3.5) {$\leftarrow$ Offload};

            \fill[blue!40] (1.5,0) rectangle (2.1, 1);
            \draw[black] (1.5,0) rectangle (2.1, 2);
            \fill[pattern=north east lines, pattern color=gray] (1.5,1) rectangle (2.1, 2);
            
            \fill[blue!40] (2.5,0) rectangle (3.1, 1);
            \draw[black] (2.5,0) rectangle (3.1, 2);
            \fill[pattern=north east lines, pattern color=gray] (2.5,1) rectangle (3.1, 2);
            
            \node[font=\tiny, text=black] at (3.8, 1.8) {Waste};
            \node[align=center, font=\scriptsize] at (1.75, -1) {(b) Volatile (Naive)};
        \end{tikzpicture}
    \end{subfigure}
    \hfill
    \begin{subfigure}[b]{0.32\linewidth}
    \centering
    \begin{tikzpicture}[scale=0.45]
        \draw[->, thick] (0,0) -- (3.5,0) node[right] {\tiny Time};
        \draw[->, thick] (0,0) -- (0,4.5) node[above] {\tiny Load};
        
        \draw[red, dashed, thick] (0, 1) -- (3.2, 1);
        \node[red, anchor=south west] at (1.1, 0.9) {\tiny VM Capacity};

        \fill[blue!40] (0.5,0) rectangle (1.1, 1);
        \draw[black] (0.5,0) rectangle (1.1, 1);
        
        \fill[orange!30] (0.5,1) rectangle (1.1, 4); 
        \fill[orange!60] (0.5,1) rectangle (1.1, 2);
        \draw[black] (0.5,1) rectangle (1.1, 4);
        
        \node[font=\tiny, anchor=west] at (1.2, 3.2) {$\leftarrow$ Old Offload};
        \node[font=\tiny, anchor=west] at (1.2, 1.8) {$\leftarrow$ Additional Offload};

        \fill[blue!40] (1.5,0) rectangle (2.1, 1);
        \draw[black] (1.5,0) rectangle (2.1, 1);
        
        \fill[blue!40] (2.5,0) rectangle (3.1, 1);
        \draw[black] (2.5,0) rectangle (3.1, 1);
        
        \node[align=center, font=\scriptsize] at (1.75, -1) {(c) Volatile (Optimal)};
    \end{tikzpicture}
\end{subfigure}
    
    \caption{Cost minimization strategies. (a) Stable load allows perfect VM matching. (b) Provisioning for \textit{average} volume under volatile load creates under-utilization (hatched) during lows. (c) Optimal allocation provisions VMs for the \textit{base} load to eliminate waste, using serverless (orange) for the peaks.}
    \label{fig:optimal_allocation}
\end{figure}

To illustrate why total load volume is an insufficient metric for cost optimization, consider the scenarios in \cref{fig:optimal_allocation}. Scenario (a) is characterized by a load that does not vary in time. Under such load, a VM provisioned at the average request rate achieves 100\% utilization and has no need for serverless. In scenarios (b) and (c), the load varies in time but the total number of requests is identical to scenario (a). Scenario (b) shows that applying that same "average-based" provisioning as used in Scenario (a) is sub-optimal. During the initial burst, the load exceeds VM capacity, forcing the rest to serverless. During the subsequent lows, the provisioned VMs sit idle, which is wasteful cost-wise. Scenario (c) shows a cost-optimal strategy, which is to provision fewer VMs to maximize their utilization and rely on serverless strictly for the bursts. 

While the example above is simplistic (in practice, sending too much traffic to FaaS merely for the sake of maximizing VM utilization can be cost-inefficient), it helps intuit Insight 3: the optimizer must account for the distribution of the load, not just its volume.

\section{\coolName{}}
\label{sec:design}

\subsection{Overview}

We introduce \coolName{}, a system that leverages the principles and insights above to reconcile strict SLO latency objectives with extreme cost efficiency. 
\coolName{} uses conventional VMs to cost-efficiently serve the bulk of the load while dynamically steering individual requests that are likely to violate SLO to serverless instances. A key challenge addressed by \coolName{} is how to split the incoming requests between VMs and elastic compute in real time so as to minimize cost by avoiding VM overprovisioning while meeting SLO targets. 

To address this challenge, \coolName{} operates at two levels of granularity. At the instance level (pod or VM), 
\coolName{} decides on a per-request basis whether the instance can serve the request within the latency budget or if the level of queueing is such that an SLO violation is likely. In the latter case, the request is redirected to a stock serverless plane (e.g., AWS Lambda). At coarse grain, \coolName{} monitors the fluctuating load across the VM instance pool and periodically adjusts the VM allocation so as to maximize overall cost-efficiency taking into account both VM and serverless costs. The two-level architecture allows \coolName{} to be both SLO-compliant despite fluctuating load and economically efficient over longer time scales.

{\em By design, \coolName{} separates the concerns of SLO enforcement from cost optimization}. The former happens at the finest grain (instance-level, on-demand, per-request); the latter is performed at coarse grain (periodically, across the entire instance pool). This explicit separation of concerns is at the heart of \coolName's effectiveness, and differentiates it from prior works, such as Libra~\cite{libra}, which conflate the two.

\begin{figure}[t]
    \centering
    
        \includegraphics[width=0.45\textwidth]{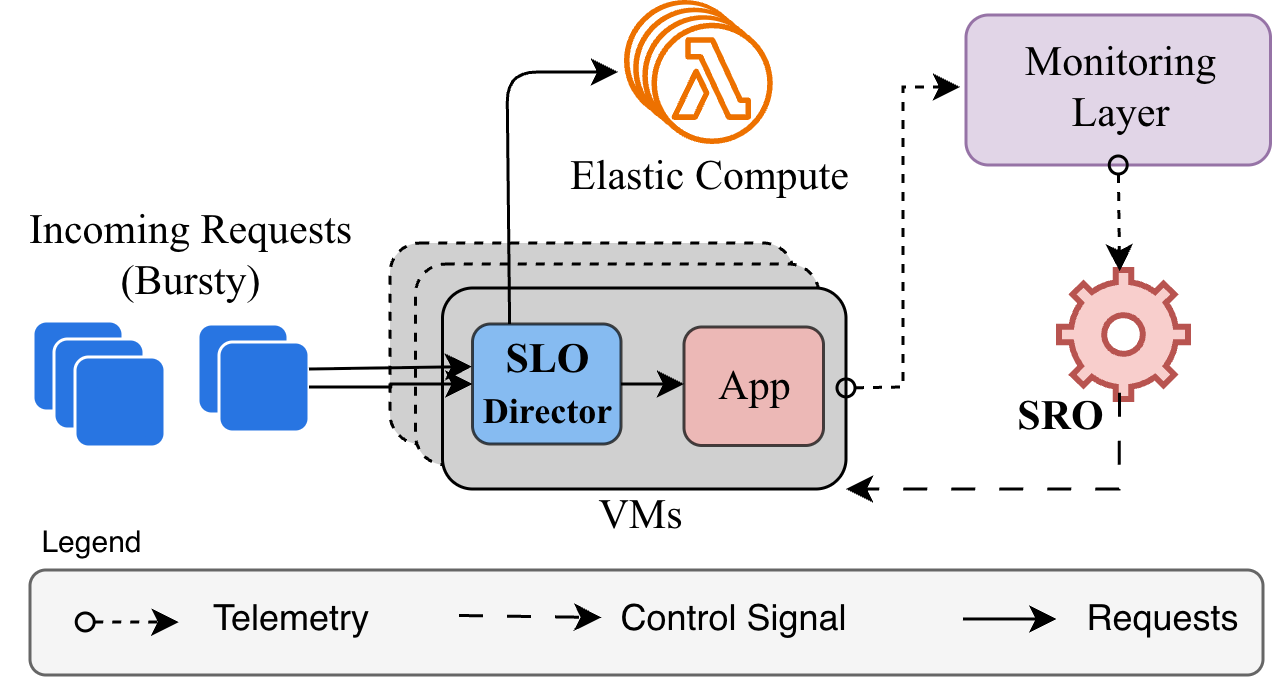}
        \caption{High-level architecture of \coolName{}}
        \label{fig:high-level-overview}
\end{figure}

\cref{fig:high-level-overview} illustrates the architecture of \coolName{}. At the instance level, a \slb{} is colocated with each application instance running on a VM. The \slb{} intercepts all incoming requests and makes a real-time decision to either serve the request locally (i.e., on the VM) or offload it to elastic compute. At the cluster level, the centralized Spandana Resource Optimizer (SRO) collects telemetry data from all instances via a monitoring layer. By analyzing the load across all instances and how effectively it is served across the two compute types, the SRO adjusts the number of provisioned VMs to optimize for cost. We next describe each component in detail.

\subsection{\slb}

The SLO Director acts as a distributed, per-instance decision point that determines whether each incoming request should be served by the colocated VM-based application container instance or offloaded to the elastic compute. The \slb{} shields the application instance from bursty demand by maintaining a bounded queue in front of the instance; when the queue is full, newly arriving requests are immediately offloaded to the elastic compute. 
All incoming requests are intercepted and passed through the \slb{}, enabling timely routing decisions without requiring any changes to clients or the application.

\slb{}'s queue acts like a leaky bucket, shaping the traffic to the local instance to avoid overloading it (details in \cref{sec:implementation:slb}) with overflow redirected to serverless. Its bounded length facilitates service time estimation, which is needed to determine if an incoming packet can meet its SLO target; given the queue length of $L_{Queue}$ and an average service rate of $\mu$, the expected queueing time is ${L_{Queue}/\mu}$.

\slb{}'s organization is informed by principles P1 and P2 (\cref{sec:bridge}, which dictate that the default service point for all requests is a VM due to its cost-efficiency, unless SLO is in jeopardy, in which case the at-risk requests (and only such requests) should be served by serverless instances. \slb{}'s specific design is derived from Insights 1 and 2 (\cref{sec:bridge}), which argue for per-request decision making and the need for queueing in front of the instance to absorb small bursts without either violating SLO or using costly serverless compute.

\subsection{\coolName{} Resource Optimizer (SRO)}

\begin{figure}[t]
    \centering
  \includegraphics[width=0.35\textwidth,trim=0 0.5cm 0 2cm,clip]{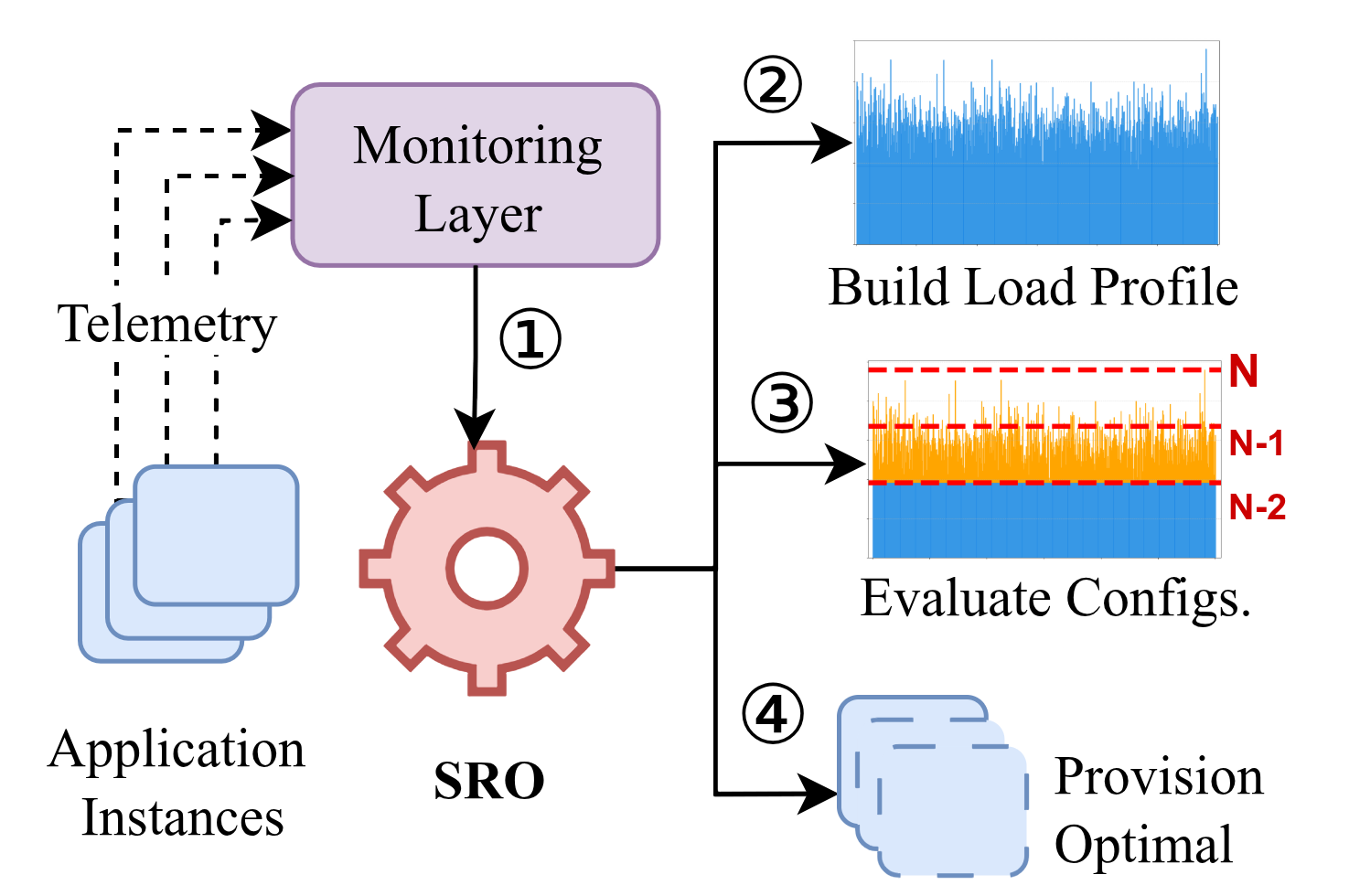}
  \setlength{\abovecaptionskip}{4pt}
        \caption{\coolName{} Resource Optimizer (SRO) in action}
        \label{fig:global_opt}
\end{figure}
While the \slb~ manages traffic in real time on individual instances, the \textit{\coolName{} Resource Optimizer (SRO)} runs as a control loop for longer-term cluster-wide optimization aimed at maintaining high utilization and cost efficiency. Existing autoscalers typically determine cluster size based on aggregated load metrics over a recent window, scaling resources only when utilization crosses a predefined threshold. Such an approach fails to account for the fine-grained volatility of the workload. The SRO differentiates itself from these traditional mechanisms by seeking a global lowest cost configuration rather than simply reacting to load volume. As detailed in Insight 3 (\cref{sec:bridge}), the SRO explicitly accounts for traffic fluctuation profile to find the optimal balance point.

To identify the optimal provisioning level, the SRO performs a "what-if" analysis by simulating various provisioning configurations against the recent high-resolution load profile. For every potential cluster size, the SRO calculates the total projected cost, which combines the fixed cost of the provisioned VMs with the estimated cost of offloading the residual bursty traffic to FaaS.

Its core logic, formalized in \cref{alg:optimizer}, is a continuous control loop triggered at a configurable coarse-grained (minutes-scale) interval.
The algorithm proceeds as follows. First, the optimizer fetches the recent load history to build a load profile (line 2). Next, it proceeds to the main for-loop (lines 7-20), which implements the core "what-if" analysis. There, the optimizer simulates a range of provisioning levels by iterating through all possible instance counts, from one instance up to the maximum required to handle the historical peak load. For each simulated instance count (inst), it first calculates the fixed cost of the VMs (line 8). Next, to determine the offload portion, it calculates the residual load by iterating through the \textit{load\_profile} and subtracting the VM cluster's capacity at each time step (lines 11-13). The cost of handling this residual load on serverless is then calculated (line 14).
After summing the VM and serverless costs to get a total projected cost, the algorithm identifies the configuration with the minimum cost (lines 16-19). Finally, if the optimal instance count \textit{(inst\_opt)} differs from the current number of running instances, the optimizer scales the number of replicas of the target application's deployment (lines 21-23).

\cref{fig:global_opt} provides an overview of the SRO’s control loop. \circled{1} The SRO begins each interval by fetching recent request-rate histories of application instances via the monitoring layer. \circled{2} Then it aggregates request-rate histories from the instances to construct a load profile (request rate vs. time), giving SRO a global view of the demand seen by the application. \circled{3} Using this profile, the SRO evaluates a range of configurations, each defined by a different number of VM instances. Each configuration corresponds to some split between the stable \textit{Base portion} handled by the VMs and the \textit{Offload portion} offloaded to the elastic compute. For each choice, the SRO calculates the combined cost of running the VMs and offloading the offload portion.  \circled{4} The SRO then identifies and provisions the lowest-cost configuration\footnote{
  Note that SRO only provisions the VMs. The offload portion is dynamically redirected to elastic compute by \slb{}s. However, the cost of elastic compute is accounted for by the SRO.}. 
This process repeats at each interval, ensuring that longer-term workload variations are accommodated by the VM provisioning, while short-term bursts are handled via the \slb{}s.

\subsection{Discussion}
\label{sec:implementation:discuss}

\beginbsec{Centralized vs. Distributed Implementation} The architecture of Spandana supports both centralized and distributed implementations. A centralized approach, which we evaluate as Spandana-C in \cref{sec:evaluation}, employs a global load balancer that routes traffic based on the aggregate state of all instance queues. Such a design benefits from global visibility, which allows the balancer to direct requests to the least-loaded instances~\cite{youtube}. However, the centralized approach requires a sophisticated load balancer tracking the state of the VMs across the deployment. The load balancer must also be present at each hop in a multi-service chain. 
\begin{algorithm}[!t]
\small
\caption{Resource Optimizer Logic}\label{alg:optimizer}
\begin{algorithmic}[1]
\Require
\Statex \(n_{\text{curr}}\): current number of application instances
\Statex \(C_{\text{vm}}\): cost parameters for VMs
\Statex \(C_{\text{serverless}}\): cost parameters for serverless backend
\Statex \(RPS_{\text{max}}\): max requests per second per instance

\Function{OptimizeResources}{}
    \State lp $\gets$ FetchLoadProfile()
    \State rps\_peak $\gets$ GetPeakRPS(lp)

    \State inst\_max $\gets \lceil rps\_peak / RPS_{\text{max}} \rceil$
    \State cost\_min $\gets \infty$
    \State inst\_opt $\gets n_{\text{curr}}$
    \Statex

    \For{$inst \gets 1$ \textbf{to} inst\_max}
    
        \State vm\_cost $\gets$ VMCost(inst, \(C_{\text{vm}}\))

        \State residual $\gets$ []
        \State vm\_capacity $\gets inst \times RPS_{\text{max}}$
        \For{each time step $t$ in $lp$}
            \State residual[$t$] $\gets \max(0, lp[t] - vm\_capacity)$
        \EndFor

        \State sl\_cost $\gets$ ServerlessCost(residual, \(C_{\text{serverless}}\))
        \State total $\gets$ vm\_cost + sl\_cost

        \If{total $<$ cost\_min}
            \State cost\_min $\gets$ total
            \State inst\_opt $\gets$ inst
        \EndIf
    \EndFor
    \Statex

    \If{inst\_opt $\neq n_{\text{curr}}$}
        \State ScaleDeployment(inst\_opt)
    \EndIf
\EndFunction
\end{algorithmic}
\end{algorithm}

Conversely, the distributed design places an SLO Director alongside each application instance to make local offloading decisions. 
We select the distributed architecture as the default implementation for Spandana. The decentralized approach aligns naturally with modern cloud-native application deployments, allowing Spandana to integrate seamlessly into existing deployments without requiring changes to the cluster-wide ingress or load balancing infrastructure. \\
\beginbsec{Feasibility of Serverless Offloading} \coolName{} maintains SLO compliance for offloaded traffic by leveraging the fact that modern FaaS latencies comfortably fit within the application's SLO target. Formally, the feasibility of offloading is governed by the inequality:$$T_{redirect} + T_{cold} + T_{exec} < \text{SLO}$$ In practice, cold starts are rare (as shown in \cref{sec:eval:main:serverless}) because frequent offloading due to load volatility naturally keeps function instances warm. Furthermore, SLO slack (commonly 5-10x of the typical service time) provides a sufficient buffer to absorb both the minimal redirection overhead and the standard execution time in the serverless tier, ensuring robust compliance.

\beginbsec{Generalizability of Compute Resources} While this work evaluates \coolName{} using a homogeneous pool of standard VMs and serverless functions as the elastic tier, the underlying architecture is agnostic to specific compute instance types. \coolName{} can readily adapt to alternative configurations, such as employing burstable instances as the elastic resource or orchestrating a heterogeneous cluster of mixed VM types to optimize steady-state costs. While incorporating these resource-level optimizations could further enhance cost-efficiency, they represent orthogonal improvements that do not alter the fundamental design principles or the key insights presented in this work.
\section{\coolName{} Under the Hood}
\label{sec:implementation}

\subsection{Technology Stack}
\beginbsec{Container Orchestration} \coolName{} is implemented on top of k8s, which manages the deployment and lifecycle of the provisioned service instances. In k8s, the fundamental unit of deployment is a pod, which is a logical group of one or more containers that share storage and network resources. In our setup, each pod contains the main application container and the sidecar container similar to how most online services are deployed today \cite{cncf2022servicemesh, dissecting}. This co-location allows the two containers to be treated as a single, scalable unit.

\beginbsec{Elastic Compute} Our current implementation uses AWS Lambda Functions as the elastic compute. Lambdas have the desirable features of fast start time (well under 1s~\cite{ustiugov:analyzing, maxday_lambda_perf}), high scalability and pay-per-use billing, all of which make it well-suited for handling bursty components of load. Importantly, \coolName{}'s design is not tied to Lambda; any future cloud service that offers fast scaling and pay-per-use billing could be used instead.

\beginbsec{Monitoring} Our monitoring layer uses a standard cloud-native stack to provide metrics for the Resource Optimizer. We use Fluent Bit \cite{fluentbit} as the log collection agent that runs on every cluster node. These agents gather logs from all sidecars and forward them to Loki \cite{loki}, a centralized aggregation system. This pipeline enables the Resource Optimizer to query Loki for RPS metrics and construct the application's time-series load profile.
\subsection{Implementing the \slb}
\label{sec:implementation:slb}

To realize the \slb, we built a custom sidecar proxy that is deployed with each application pod. The proxy is written in Go and mimics the core functionality of production-grade proxies such as Envoy \cite{envoy}, namely traffic interception and forwarding. A custom proxy allowed us to embed \slb's logic without the challenge of modifying a more complex codebase. This is not a limitation, however, and another implementation could instead deploy \slb{} logic as part of an existing sidecar.

\beginbsec{Queueing and Offloading Logic} The \slb{}, residing in a sidecar, maintains a bounded FIFO request queue \textit{(RQ)} to buffer incoming load. The size of the RQ is configurable and sets the maximum local backlog tolerated before offloading. Any request that arrives when the RQ is at this maximum capacity is immediately offloaded by invoking an AWS Lambda function with the request payload. For offloaded requests, \slb{} subsequently receives the response and forwards it back to the client over the original connection. The offloading process is completely opaque to the client requiring no client-side changes.

\beginbsec{Request Pacing and SLO Adherence}
To create a smooth load on the local application instance, the \slb~ implements a leaky bucket-like pacing mechanism. It spawns a dedicated thread that dequeues and forwards requests to the local instance. The time between forwarded requests is governed by a \textit{pacing interval}, calculated as  \(T_{\text{wait}} = 1 / RPS_{\text{max}}\), where  ($RPS_{\text{max}}$) is the application instance's maximum sustainable request handling capacity per-second, determined via profiling.

The controller continuously monitors the instance's response latency. If the response latency exceeds the SLO target of the deployed service, the controller dynamically increases the pacing interval (thus lowering the rate) to alleviate pressure on the application. In practice, we have found that the ability of the controller to adjust the pacing rate for its individual instance (rather than using a fixed value for all instances) is helpful because actual performance of different instances of the same type varies across the deployment. By adjusting its pacing rate, each instance can maximize throughput in an SLO-compliant manner.

\beginbsec{Metric Exposure} 
To support cluster-level optimization by the SRO, the \slb~ records request-level metadata within each pod. Specifically, it logs arrival timestamps, response latencies, and whether each request was served locally or offloaded. These logs are written in a structured format such that they can be consumed by the monitoring layer without interfering with the application.

\subsection{Implementing the SRO}

Spandana Resource Optimizer (SRO) is implemented as a standalone \textbf{Go application} deployed as a standard k8s Deployment. This design ensures the SRO runs continuously alongside the application workload, leveraging k8s' native resilience features. \\ \textbf{Data Ingestion and Profiling:} To build the load profile required for optimization, the SRO interfaces with \textbf{Loki}, our centralized log aggregation system. At the start of every control interval (configured to 2 minutes), the SRO queries Loki to fetch aggregated RPS metrics from all active SLO Directors. We utilize Loki's LogQL query language to efficiently retrieve and build the load profile from the preceding two minutes of traffic. \Cref{app:sro-history-window} evaluates the effect of the load history window on deployment cost.\\ \textbf{Control Loop Execution:} The optimization loop runs in a dedicated goroutine triggered by a ticker. To ensure stability, we employ a configurable cooldown period between scaling actions. In our measurements, the optimizer consumes less than 1\% of a single vCPU with a negligible memory footprint, making it lightweight.

\subsection{\coolName{} in a Microservice Chain}

\coolName{}'s design extends naturally to applications composed of microservice chains. 
When an application container running on a VM handles a request and subsequently calls a downstream microservice, \coolName{} allows the request to follow the standard intra-cluster communication path. The request is routed using k8s' internal service discovery mechanisms \cite{kubernetes_service_discovery} to an instance of the appropriate downstream microservice in the cluster. The \slb{} at the destination instance then intercepts the call and applies its queueing and offloading logic. 

Once the serverless instance completes processing, the request is routed back to the VM cluster by configuring the elastic compute to direct downstream requests to the k8s service endpoint of the target microservice. Upon arrival, the request is intercepted by the \slb~ at the downstream microservice, which manages the request  like any other incoming request. This approach ensures that each request at {\em every} hop in the microservice chain has an opportunity to run on a VM, hence ensuring high VM utilization and low cost, while retaining the ability to execute on elastic compute in order to meet SLO if the VM instance is at capacity.

\subsection{Portability and Developer Effort}
\coolName{} does not assume that all VM-hosted services can run unchanged on a FaaS platform. Porting to serverless is most straightforward for stateless services (or services whose state is already externalized to managed storage). Services with their own persistent state are treated as VM-only in Spandana. For stateless services, our experience shows that the porting effort is minimal. Serverless platforms such as AWS Lambda provide native support for container images \cite{aws-lambda-container}, allowing the same container to run on both k8s (VM-backed) clusters and Lambda without repackaging. For applications initially designed for VMs, tools such as the AWS Lambda Web Adapter \cite{aws-lambda-web-adapter} and Serverless Adapter \cite{serverless-adapter} translate Lambda's event-based model into standard HTTP requests, enabling applications to execute in Lambda without modifications.
Any service that is difficult to port need not be ported; Spandana can be viewed as a best-effort optimization from the developer's standpoint. In a microservice graph, some services would then be "Spandana-fied" to reduce cost under SLO constraints, others would run as VM-only.

\section{Experimental Methodology}
\label{sec:methodology}

\beginbsec{Studied Applications}
We evaluate \coolName{} on four cloud applications that span popular language runtimes and workload characteristics. The Ratings and Details web services are drawn from the BookInfo application \cite{bookinfo}; they are I/O-bound services implemented in Node.js and Ruby, respectively. The other two, Image Processing and Compression, are CPU-bound data-processing Python applications adapted from SeBS \cite{sebs}. All applications are configured to use managed cloud databases, reflecting common cloud deployment practices. Single-request latencies on an idle instance are 7ms (Details), 14ms (Ratings), 40ms (Image Processing), and 130ms (Compression). 

\begin{figure}[t]
    \centering
    \includegraphics[width=0.43\textwidth]{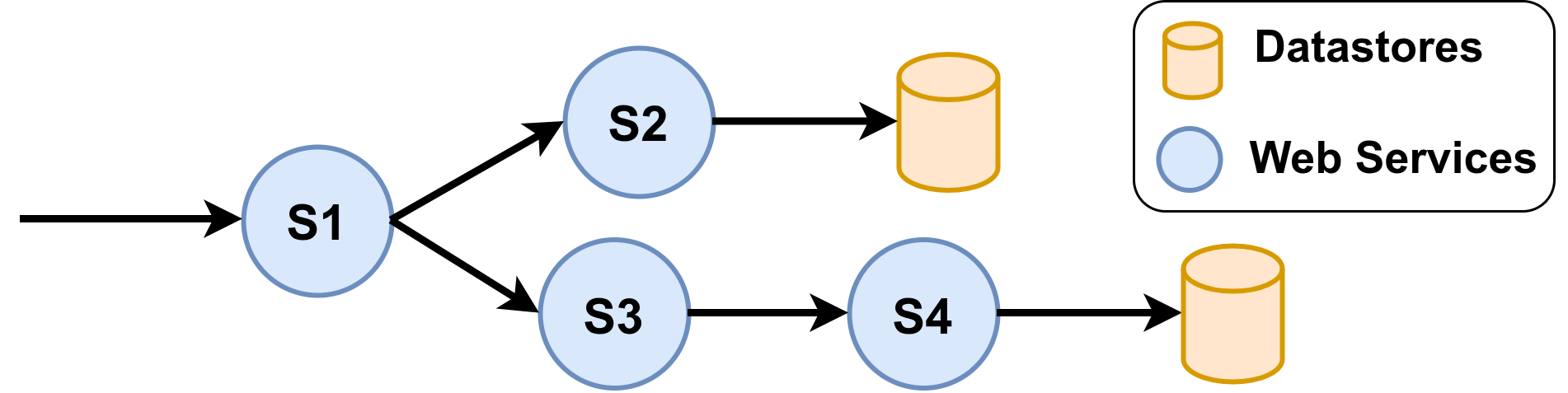}
    \caption{Service topology of BookInfo application}
    \label{fig:chains-topology}
\end{figure}

We also evaluate an application with a chain of microservices. For this, we use the BookInfo application, which consists of four web services — Productpage (S1), Details (S2), Reviews (S3), and Ratings (S4) — arranged in a mix of sequential and fan-out dependencies, as shown in \cref{fig:chains-topology}. The leaf services in this chain are configured to use DynamoDB as their datastore backend. The SLO is defined with respect to the entire chain and is set to 10x the latency of a single request traversing an otherwise idle chain.

\beginbsec{Load Trace \& Replay Setup}
For our main experiments, we use a one-hour Twitter trace \cite{twitter:trace} to drive a realistic load on a cloud service. Other public traces report load at even coarser granularity~\cite{serverless-wild} hence obscuring fine-grained behavior, or are much older \cite{wikibench_project}. The chosen trace exhibits a steady load at coarse granularity with fine-grained fluctuations as discussed in \cref{sec:motivation:bursts}, making it representative of an online service in steady state.
We scaled up the trace \cite{traceupscaler} to generate sufficient load for the studied applications to run on a multi-node cluster with at least 10 instances per application. 
We replay the load trace on each application three times and report the run with the median SLO violation rate.
To evaluate \coolName{}’s ability to handle unexpected load spikes, we selected a different one-hour Twitter trace that includes a load surge peaking at ~2× the mean load. We use k6 \cite{k6}, a modern load generator, to replay the trace. K6 runs on a dedicated VM (4 vCPUs, 16 GB of RAM) in the same AWS region as our k8s cluster.

\vspace{2mm}
\beginbsec{Studied Systems}

\textbf{1) Kubernetes HPA} - As a baseline representing common industry practice, we use HPA, the default autoscaling mechanism in k8s. The HPA is a simple yet widely-used autoscaling mechanism, with recent industry reports highlighting its popularity among organizations running k8s~\cite{datadog_container_report}. We study two configurations of HPA:
\\
\textbf{HPA-S (Standard)}: Configured to scale based on CPU utilization with a target threshold of 60\%, a common production setting to balance performance and cost\cite{gke_autoscale_docs, madu}. \\
\textbf{HPA-O (Oracle)}: Configured to just meet the latency SLO (using a-priori knowledge of the entire trace), thus resulting in the most cost-efficient SLO-compliant HPA setup possible.

\textbf{2) Libra} - We compare against Libra \cite{libra}, a state-of-the-art hybrid framework that is the closest related work to \coolName{}. Libra attempts to balance cost and SLO by proactively engaging FaaS alongside VMs. To ensure a fair comparison under high-throughput conditions, we implemented a high-performance version of Libra in Go. Consistent with our own setup, we use m5a.large EC2 VM instances for the IaaS tier and AWS Lambda for the FaaS component in our Libra implementation. We set its key parameter $p$ 
to 80\%, a value suggested by the original authors and empirically verified in our setup as the threshold that just meets the SLO target.

\textbf{3) AutoBurst} - We compare against AutoBurst \cite{autoburst_socc}, a SOTA research system that combines hybrid compute types in the form of low-cost burstable\footnote{Burstable VMs are intended for low average CPU utilization (e.g., 20\%)~\cite{aws_ec2_burstable_credits} but can burst to higher utilization on-demand. When utilization is below a threshold, burstable instances accumulate credits that can then be spent in high-utilization periods.} instances and regular on-demand VMs. Stated goals for AutoBurst are to minimize cost while meeting SLO.

AutoBurst requires configuring several parameters, notably the initial numbers of on-demand and burstable instances at experiment start, and the \textit{desired credit level} for burstables.
Guided by communication with the work's authors, we size the on-demand pool to sustain the average load and configure the burstable pool to absorb demand beyond the provisioned on-demand capacity.
We set the \textit{desired credit level} based on the load fluctuations of our trace and include a warm-up period for burstable instances to accumulate credits to this level before experiments begin. 
For PD controller parameters, we adopt the defaults suggested by AutoBurst. 
Consistent with the original AutoBurst paper, we use \textit{m5a.large} instances as the on-demand compute (same as what \coolName{} uses for its VMs) and \textit{t3a.small} instances as the low-cost, burstable compute.

\textbf{4) \coolName{}} - Default is a decentralized design where the \slb~ is co-located with each VM to make local, per-request offloading decisions. 
We empirically tuned two parameters of \coolName{}: the \slb{} queue length and the SRO control loop interval. For \slb{}, the queue length must absorb short bursts on the VM without adding queueing delay that violates the latency SLO. For leaf services, we set the queue length to nine requests, allowing up to nine waiting requests while keeping the requests within the $10\times$ latency SLO when served locally on the VM. For non-leaf services, we double the queue length to 18, because time spent waiting on calls to downstream services allows a larger backlog before CPU saturation. More generally, the queue length should be set per service based on its concurrency tolerance, such as I/O overlap or multi-threading and latency SLO target. \Cref{app:slo-director-queue-length} examines the sensitivity of cost, offloading, and SLO compliance to the queue length. For SRO, the control loop runs every two minutes, an interval chosen to account for the startup latency of new VMs and avoid scaling oscillations. This choice aligns with common autoscaling practice, where minute-scale intervals balance responsiveness
and stability~\cite{Autopilot, madu}.

We also evaluate a centralized variant of \coolName{}, denoted as \textbf{Spandana-C}. This variant uses a centralized load balancer at the cluster's entry point that maintains a global view of all individual instance queues. While such an omniscient load balancer is impractical in large-scale production clusters due to the overhead and complexity of maintaining real-time global state of all VMs, it serves as a useful "near-optimal" baseline for load balancing across all application instances. By forwarding traffic to the least-loaded instance using its perfect knowledge, Spandana-C allows us to isolate the cost and performance benefits of having global visibility for request placement, thus reducing the need for serverless through the ability to choose the least-loaded VM. Similar to the primary decentralized architecture, Spandana-C still leverages serverless offloading: if the load balancer detects that the queues on all active VMs are saturated, it routes incoming requests directly to FaaS. 

\beginbsec{Cluster} Our experiments were run on an AWS EKS cluster using m5a.large EC2 VM instances. For the elastic compute, we use AWS Lambda. 
Note that we do not rely on pre-warmed functions or attempt to keep functions warm using keep-alive (dummy) requests or similar mechanisms. Thus, all presented results intrinsically capture various Lambda-related artifacts including cold starts.

\beginbsec{Evaluation Metrics} We compare \coolName{} against the baselines using CPU utilization, SLO adherence, and cost. 
Each application instance is provisioned with 1 vCPU, and we report the CPU utilization of the main application container to ensure a fair comparison across systems. 
For SLO, we define the latency target as 10x the latency of a single request on an idle application instance and adopt a strict SLO violation limit of 1\%, meaning that at most 1\% of requests may exceed the target latency. 
Finally, we report monetary cost based on AWS pricing as of September 2025 covering both VM instances (AWS EC2) and serverless invocations (AWS Lambda).

\section{Evaluation}
\label{sec:evaluation}

\subsection{Main Results: SLO/Cost/Utilization}
\label{sec:evaluation:util-slo}

We first evaluate \coolName{}’s ability to improve CPU utilization and reduce cost while adhering to strict SLO. We start with the latter, a firm requirement that must be met for good user experience.

{\bf SLO:} \cref{tab:slo_violations} shows the SLO violation rates for all systems across the four applications. Both \coolName{} variants and both HPA variants meet the strict 1\% SLO violation budget on all applications. Spandana-C achieves the lowest SLO violation rate of all studied systems as it uses its omniscient load balancer to place requests and quickly redirects to serverless if all instances experience queueing. As expected, HPA-O has a higher violation rate than HPA-S since the former is optimized for cost under SLO compliance using oracle knowledge of the trace. HPA-S must be overprovisioned to contend with unknown load, resulting in fewer violations but higher cost as shown below. AutoBurst and Libra exceed the 1\% SLO budget for \textit{Details service}. The Details service presents the greatest SLO challenge among the evaluated applications due to its low typical service time, resulting in the lowest SLO target of just 70ms and making it highly sensitive to even minor delays. For other applications, both AutoBurst and Libra meet the SLO target but with higher violation rates than either HPA or \coolName. The reason for AutoBurst and Libra having higher violation rates is their inability to handle fine-grained bursts. Lacking a mechanism to offload traffic at per-request granularity, they continue routing fine-grained bursts to already-overloaded VMs, directly causing the queueing delays that result in SLO violations.

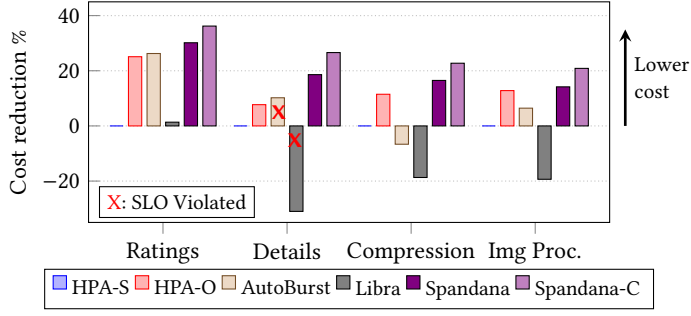
\begin{figure}[t!]
\centering
\begin{tikzpicture}
\begin{axis}[
  ybar,
  bar width=5pt,
  width=\linewidth,
  height=4cm,
  ymin=-35,
  ymax=41,
  enlarge x limits=0.20,
  ylabel={Cost reduction \%},
  xticklabel style={align=center},
  symbolic x coords={Ratings,Details,Compression,{Img Proc.}},
  xtick=data,
  xtick pos=left,  
    ytick pos=left, 
  ymajorgrids,
  grid style={densely dotted},
  legend style={at={(0.5,-0.21)}, anchor=north, legend columns=6,font=\small},
  legend image code/.code={
      \draw[#1, /pgfplots/ybar legend]
      (0cm,0cm) rectangle (5pt,5pt);
  },
  clip=false,
]

\addplot+[ybar] coordinates {
    (Ratings, 0) 
    (Details, 0) 
    (Compression, 0) 
    (Img Proc., 0)
};

\addplot+[ybar] coordinates {
    (Ratings, 25.1) 
    (Details, 7.71) 
    (Compression, 11.484) 
    (Img Proc., 12.81)
};

\addplot+[ybar] coordinates {
    (Ratings, 26.26) 
    (Details, 10.21) 
    (Compression, -6.64) 
    (Img Proc., 6.44)
};


\addplot+[ybar] coordinates {
    (Ratings, 1.36) 
    (Details, -30.997) 
    (Compression, -18.7) 
    (Img Proc., -19.35)
};

\addplot+[ybar,violet,draw=black] coordinates {
    (Ratings, 30.16) 
    (Details, 18.6) 
    (Compression, 16.5) 
    (Img Proc., 14.17)
};

\addplot+[ybar,violet!50,draw=black] coordinates {
    (Ratings, 36.21) 
    (Details, 26.6) 
    (Compression, 22.74) 
    (Img Proc., 20.87)
};

 \node[anchor=center, font=\sffamily\small, text=red,xshift=-3pt] 
    at (axis cs:Details,5) {\textbf{X}};

 \node[anchor=center, font=\sffamily\small, text=red,xshift=3pt] 
    at (axis cs:Details,-5) {\textbf{X}};

    \draw [->, >=stealth, line width=1pt] 
    ({rel axis cs:1.03,0} |- {axis cs:Ratings,0}) -- ({rel axis cs:1.03,0} |- {axis cs:Ratings,35}) 
    node [midway, right, align=left, font=\small] {Lower\\cost};
\legend{HPA-S,HPA-O, AutoBurst, Libra, Spandana, Spandana-C}
\node[draw, fill=white, anchor=south west, font=\small] 
    at (rel axis cs: 0.02, 0.02) {\textcolor{red}{X}: SLO Violated};
\end{axis}
\end{tikzpicture}
\caption{Cost reduction compared to HPA-S (higher is better). 
}
\label{fig:cost-all}
\end{figure}
\begin{figure}[t!]
    \centering
    \begin{tikzpicture}
    \begin{axis}[
        ybar, 
        height=4cm,
        width=0.5\textwidth,
        enlarge x limits=0.18,
        legend style={at={(0.5,-0.21)}, anchor=north, legend columns=6,font=\small},
        legend image code/.code={
        \draw[#1, /pgfplots/ybar legend]
        (0cm,0cm) rectangle (5pt,5pt);
        },
        ylabel={CPU Utilization \%},
        symbolic x coords={Ratings, Details, Compression, Img Proc.},
        xtick=data,
        xtick pos=left,  
        ytick pos=left, 
        ymin=50,
        ymax=90,
        bar width=5.2pt,
        xticklabel style={align=center}, 
    ]

    \addplot coordinates {(Ratings, 56.1) (Details, 56) (Compression, 58.9) ({Img Proc.}, 58.9)};

    \addplot coordinates {(Ratings, 74.2) (Details, 64.4) (Compression, 66.6) ({Img Proc.}, 71.3)};
        
    \addplot coordinates {(Ratings, 75.34) (Details, 76) (Compression, 54.3) ({Img Proc.}, 71.5)};

    \addplot coordinates {(Ratings, 72.8) (Details, 68.9) (Compression, 58.1) ({Img Proc.}, 61.5)};
    
    \node[anchor=center, font=\sffamily\small, text=red,xshift=-3pt] 
    at (axis cs:Details,62) {\textbf{X}};

    \node[anchor=center, font=\sffamily\small, text=red,xshift=3pt] 
    at (axis cs:Details,62) {\textbf{X}};



    \addplot[fill=violet] coordinates {(Ratings, 86.5) (Details, 78) (Compression, 76.5) ({Img Proc.}, 76.5)};

    \addplot[fill=violet!50] coordinates {(Ratings, 85.4) (Details, 76.2) (Compression, 79) ({Img Proc.}, 79.5)};

    \legend{HPA-S, HPA-O, AutoBurst, Libra, Spandana, Spandana-C}
    \node[draw, fill=white, anchor=north west, font=\small] 
    at (rel axis cs: 0.71, 0.98) {\textcolor{red}{X}: SLO Violated};

    \end{axis}
    \end{tikzpicture}
    \caption{CPU utilization (higher is better).}
    \label{fig:cpu-all}
\end{figure}
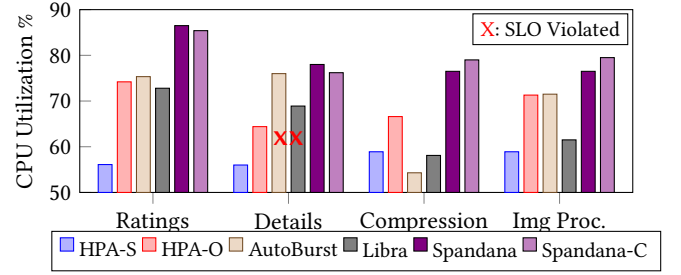

\begin{table}[t!]
\centering
\small
\setlength{\tabcolsep}{2pt}
\begin{tabular}{lcccccc}
\toprule
\textbf{App} & \textbf{\coolName{}} & \textbf{Spandana-C} & \textbf{HPA-S} & \textbf{HPA-O} & \textbf{AutoBurst} & \textbf{Libra}\\
\midrule
Ratings & 0.05 & 0.01 & 0.02 & 0.80 & 0.32 & 0.60 \\
Details & 0.34 & 0.14 & 0.14 & 0.54 & \textcolor{red}{\textbf{1.14}} & \textcolor{red}{\textbf{1.59}} \\
Compr & 0.01 & 0.01 & 0.02 & 0.57 & 0.15 & 0.07 \\
Img Proc & 0.01 & 0.00 & 0.02 & 0.74 & 0.23 & 0.20 \\
\bottomrule
\end{tabular}
\caption{SLO violations (\%). \textcolor{red}{\textbf{Red}} indicates violation of the 1\% SLO target.}
\label{tab:slo_violations}
\end{table}

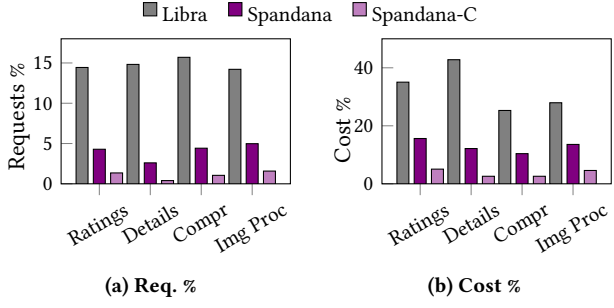
\begin{figure}[t!]
\captionsetup[subfigure]{skip=2pt}
    \centering
    \pgfplotsset{
        double_col_bar/.style={
            ybar,
            width=1.1\linewidth,
            height=3.5cm,
            enlarge x limits=0.25,
            ylabel style={align=center, yshift=-5pt},
            symbolic x coords={Ratings, Details, Compr, {Img Proc}},
            xtick=data,
            xtick pos=left,
        ytick pos=left,
            ymin=0,
            xticklabel style={align=center, font=\small, yshift=2pt},
            yticklabel style={font=\small},
        nodes near coords style={/pgf/number format/fixed, /pgf/number format/precision=1, font=\tiny}, 
            bar width=4.5pt,
            legend style={
                font=\small,
                at={(0.5,3.35)},
                anchor=north,
                legend columns=-1,
                draw=none,
                fill=none,
                /tikz/every even column/.append style={column sep=0.2cm}
            },
            legend image code/.code={
                \draw[#1, /pgfplots/ybar legend]
                (0cm,0cm) rectangle (5pt,5pt);
                },
        }
    }
    \ref{commonlegend}
    \begin{subfigure}[b]{0.495\columnwidth}
        \centering
        \begin{tikzpicture}
            \begin{axis}[
                double_col_bar,
                ylabel={Requests \%},
                ylabel style={align=center, xshift=-2pt},
                ymax=18,
                xticklabel style={rotate=30},
                legend entries={Libra, Spandana, Spandana-C},
                legend to name=commonlegend,
            ]
                \addplot[fill=black!50] coordinates {(Ratings, 14.44) (Details, 14.82) (Compr, 15.7) ({Img Proc}, 14.21)};
                \addplot[fill=violet] coordinates {(Ratings, 4.29) (Details, 2.6) (Compr, 4.42) ({Img Proc}, 4.98)};
                \addplot[fill=violet!50] coordinates {(Ratings, 1.35) (Details, 0.4) (Compr, 1.05) ({Img Proc}, 1.58)};
            \end{axis}
        \end{tikzpicture}
        \caption{Req. \%}

    \end{subfigure}
    \hfill
    \begin{subfigure}[b]{0.495\columnwidth}
        \centering
        \begin{tikzpicture}
            \begin{axis}[
                double_col_bar,
                ylabel={Cost \%},
                ylabel style={align=center, xshift=-7pt},
                xlabel style={align=center, xshift=-7pt},
                xticklabel style={rotate=30},
                ymax=50,
            ]
                \addplot[fill=black!50] coordinates {(Ratings, 35.05) (Details, 42.79) (Compr, 25.29) ({Img Proc}, 27.95)};
                \addplot[fill=violet] coordinates {(Ratings, 15.62) (Details, 12.16) (Compr, 10.4) ({Img Proc}, 13.58)};
                \addplot[fill=violet!50] coordinates {(Ratings, 5.06) (Details, 2.6) (Compr, 2.6) ({Img Proc}, 4.6)};
            \end{axis}
        \end{tikzpicture}
        \caption{Cost \%}
    \end{subfigure}

    \caption{Breakdown of FaaS request volume (a) and cost contribution (b) across applications.}
    \label{fig:serverless-cost-requests}
\end{figure}

{\bf Cost-efficiency:}  \cref{fig:cost-all} presents the cost reduction for the studied systems relative to HPA-S. Note that higher is better in the figure, while negative values indicate a cost increase relative to HPA-S. AutoBurst reduces costs for three applications by 6-26\% (though the savings on the Details service come at the cost of an SLO violation) and incurs a cost increase of $\approx$7\% for the Compression service. Libra performs poorly, increasing costs by 19-31\% on three out of four applications. We find that the inefficiency of Libra stems from sub-optimal load balancing decisions that offload a significant volume of requests to expensive FaaS resources (request \& cost distribution shown in \cref{fig:serverless-cost-requests}). In contrast, both variants of \coolName{} consistently reduce costs across all workloads while strictly adhering to SLOs. \coolName{} achieves cost reductions of 14-30\% (averaging 19.9\%), consistently outperforming even Oracle HPA (HPA-O). \coolName{}-C further optimizes efficiency, achieving reductions of 21-36\% (averaging 26.6\%) by leveraging the central load balancer's omniscient knowledge of VMs' load to make better request routing decisions.

{\bf CPU utilization:} \cref{fig:cpu-all} plots the average CPU utilization achieved by the schemes. HPA-S consistently runs at low utilization (approximately 56-59\%) due to its conservative scaling threshold. AutoBurst and Libra achieve average utilization of 54-76\% and 58-73\%, respectively. While mostly better than HPA-S, both are significantly inferior to \coolName.
Both \coolName{} variants constantly achieve the highest utilization ranging from 76-86\%  across all applications with strict SLO compliance (unlike AutoBurst and Libra). \coolName{} pushes utilization to 76.5-86.5\%, significantly outperforming HPA-S and surpassing even HPA-O on all workloads. \coolName{}-C performs comparably. Both \coolName{} variants safely operate resources near saturation without compromising SLO.

\begin{table}[t!]
\centering
\small
\begin{tabular*}{\columnwidth}{@{\extracolsep{\fill}} l c c c @{}}
\toprule
 & \textbf{\coolName{}} & \textbf{HPA-O} & \textbf{HPA-S} \\
\midrule
SLO Violations. (\%) $\downarrow$ & 0.05 & 0.45 & \textbf{0.00} \\
Total Cost (\$) $\downarrow$      & \textbf{1.89} & 1.94 & 2.29 \\
Avg. CPU Utilization (\%) $\uparrow$     & \textbf{77.0} & 70.6 & 59.3 \\
\bottomrule
\end{tabular*}
\caption{End-to-end SLO violations (\%), Total Cost (\$), and Average CPU Utilization (\%) for BookInfo service chain.}
\label{tab:chains_comparison}
\end{table}

    \begin{figure}[t!]
      \captionsetup{skip=0pt}
        \centering
        \includegraphics[width=0.45\textwidth]{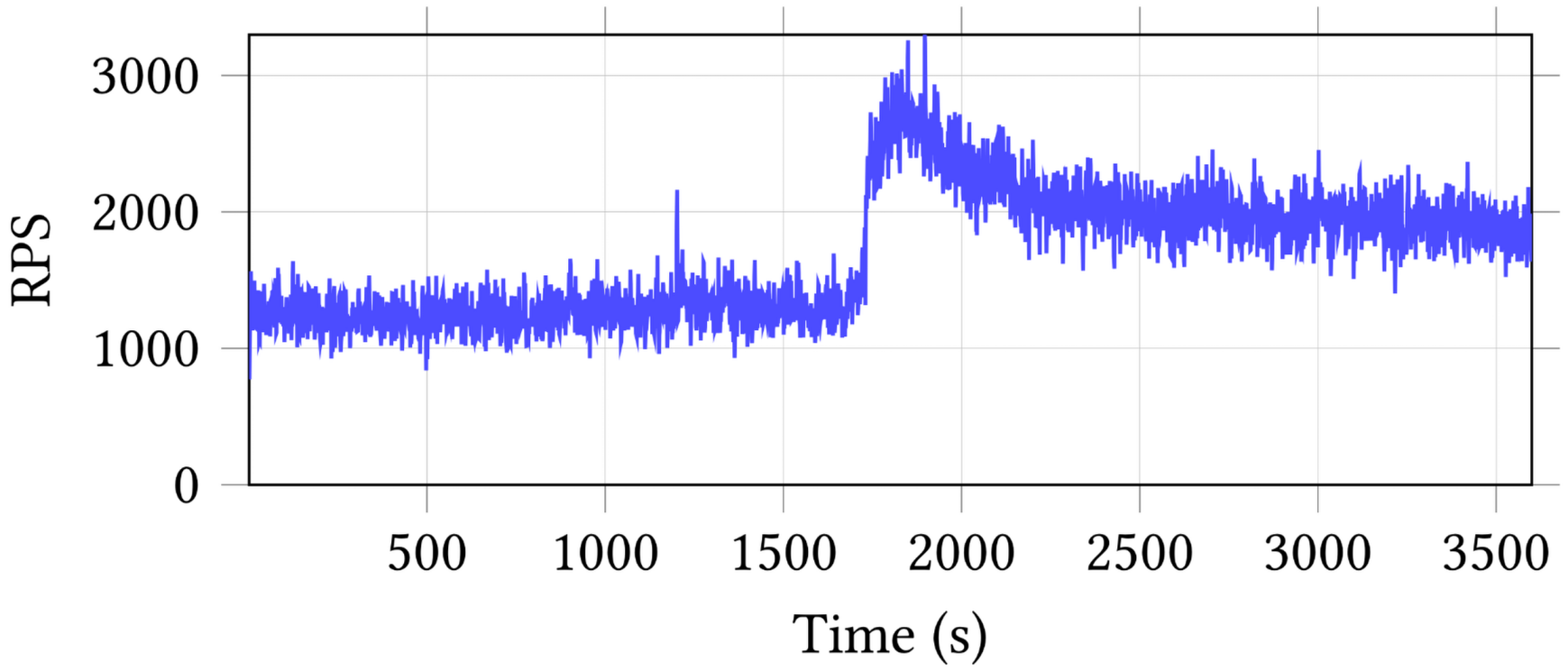}
        \setlength{\abovecaptionskip}{-1pt}
        \caption{1-hour Twitter trace with a sudden load surge.}
        \label{fig:load-spike-trace}
    \end{figure}

{\bf FaaS:}
\label{sec:eval:main:serverless}
\cref{fig:serverless-cost-requests} shows the fraction of requests served by FaaS and the corresponding cost proportion for Libra and variants of \coolName{}. Across all applications, \coolName{} offloads only a small fraction of requests, ranging from 2.6\% to 5\%, which limits the FaaS portion of the total cost to between 10\% and 15\%. \coolName{}-C lowers both numbers, offloading merely 0.4-1.6\% of requests and keeping FaaS costs below 6\% for all workloads, thanks to its omniscient request routing. In contrast, Libra consistently offloads more requests to FaaS (roughly 15\% of the total), resulting in FaaS charges consuming a large share of the overall cost at 25-43\%.

We also assess the impact of FaaS cold starts on \coolName{}. Recall that \coolName{} offloads only a tiny fraction of total traffic (2.6--5\%) to FaaS. Within this already small volume, we find that cold starts affect a negligible percentage of requests: 0.03\% for \textit{Ratings}, 0.13\% for \textit{Details}, 0.10\% for \textit{Compression}, and just 0.01\% for \textit{Image Processing}. The low cold start rate partly explains why SLO is not compromised despite cold starts in the critical path of the request latency. The other reason why cold starts do not affect SLO is in the case of compute-intensive workloads (\textit{Compression} and \textit{Image Processing}), whose longer processing times result in a higher SLO target that is large enough to absorb the overhead of a cold start.

\subsection{Service Chains}
\label{sec:evaluation:chains}

We next evaluate \coolName{} on service chains using the BookInfo application, shown in \cref{fig:chains-topology}. \coolName{} works naturally for chains because the distributed \slb{} logic applies independently at each hop and uses existing cloud infrastructure.
In contrast, \coolName{}-C, AutoBurst and Libra do not "natively" support chains, requiring significant modifications and additional costs (in the form of centralized routing and decision-making components {\em at each hop}). Thus, we compare only to HPA as the only system that, like \coolName, natively works for chains.

For this study, SLO violations are calculated from end-to-end chain latency, while cost and utilization are aggregated across the deployment.  Results are summarized in \cref{tab:chains_comparison}. \coolName{} comfortably meets the SLO target with a violation rate of just 0.05\% while delivering the lowest total cost, undercutting HPA-S by 17\% and the Oracle HPA by 2.6\%. This cost efficiency is driven by better resource usage: \coolName{} maintains an average CPU utilization of 77.0\% across all services in the chain, significantly higher than both HPA-S (59.3\%) and HPA-O (70.6\%).

These results demonstrate that \coolName{}’s benefits extend beyond individual applications to entire chains of microservices, where \coolName{} consistently increases utilization and lowers cost while remaining within the SLO budget.

\subsection{Handling Load Spikes}
\label{sec:evaluation:spikes}

    \begin{figure}[t!]
        \centering
        \begin{tikzpicture}
\begin{axis}[
    width=0.95\columnwidth,
    height=3.8cm,
    grid=both,
    grid style={line width=.1pt, draw=gray!30},
    major grid style={line width=.1pt, draw=gray!50},
    tick align=outside,
    enlargelimits=false,
    tick label style={/pgf/number format/1000 sep={}},
    xmin=1600,
    xmax=2200,
    ymin=0,
    ymax=1050,
    xlabel={Time (s)},
    ylabel={Latency (ms)},
    legend style={
        at={(0.5,-0.45)},
        anchor=north,
        legend columns=3,
        cells={anchor=west},
        font=\small,
        /tikz/every even column/.append style={column sep=10pt}
    },
    xtick={1600,1700,1800,1900,2000,2100,2200},
    ytick={0,200,400,600,800,1000},
]

\addplot[blue, very thick] table[x=time, y=latency, col sep=comma] {figures/eval/p99_latency/latency_p99_Spandana.csv};
\addlegendentry{Spandana}

\addplot[blue!50, very thick] table[x=time, y=latency, col sep=comma] {figures/eval/p99_latency/latency_p99_Spandana-C.csv};
\addlegendentry{Spandana-C}

\addplot[yellow!60!orange, very thick] table[x=time, y=latency, col sep=comma] {figures/eval/p99_latency/latency_p99_HPA.csv};
\addlegendentry{HPA-S}

\addplot[green!40!black, very thick] table[x=time, y=latency, col sep=comma] {figures/eval/p99_latency/latency_p99_AutoBurst.csv};
\addlegendentry{AutoBurst}

\addplot[violet, very thick] table[x=time, y=latency, col sep=comma] {figures/eval/p99_latency/latency_p99_Libra.csv};
\addlegendentry{Libra}

\addplot[dashed, red, very thick] coordinates {(1600,140) (2200,140)};
\addlegendentry{SLO Target}

\addplot[orange, ultra thick] coordinates {(1600,1000) (2200,1000)};
\draw[->, thick, black] 
  (axis cs:2150, 650) -- (axis cs:2150, 1000);
\node[black, anchor=north east, font=\small] at (axis cs:2200, 600) {Timeout};
\end{axis}
\end{tikzpicture}
        \setlength{\abovecaptionskip}{0pt}
        \caption{P99 latency  of \textit{Ratings} service under load spike}
        \label{fig:spike-p99}
    \end{figure}
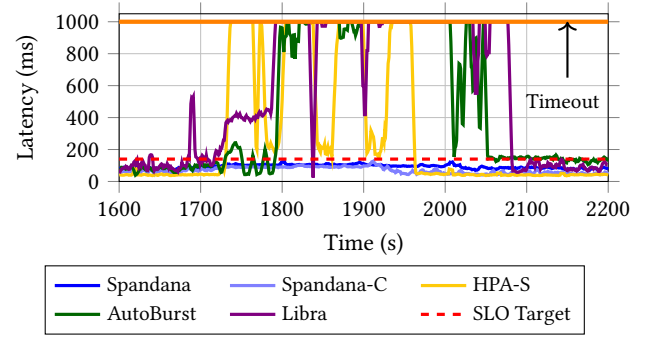

\begin{table}[t!]
\small
\centering
\setlength{\tabcolsep}{2pt}
\begin{tabular}{lccccc}
\toprule
 & \textbf{\coolName{}} & \textbf{\coolName{}-C} & \textbf{HPA-S} & \textbf{AutoBurst} & \textbf{Libra} \\
\midrule
SLO vio.(\%) $\downarrow$  & 0.08 & \textbf{0.02} & \textcolor{red}{\textbf{5.96}} & \textcolor{red}{\textbf{7.87}} & \textcolor{red}{\textbf{9.43}} \\
Cost (¢) $\downarrow$            & 74.4 & \textbf{57.5} & 81.8 & 90.8 & 95.4 \\
CPU Util. (\%) $\uparrow$ & \textbf{92.0} & 87.0 & 58.0 & 71.1 & 73.5 \\
\bottomrule
\end{tabular}
\caption{CPU utilization (\%), SLO violations (\% of requests above SLO target), and cost (¢) under a load spike. }
\label{tab:spike-summary}
\end{table}

To evaluate \coolName{} under sudden demand surges, we use a Twitter trace that includes a sharp load spike, shown in \cref{fig:load-spike-trace}. The spike arrives at $\approx$1700s and eventually settles at a higher load level than the pre-spike load level.
We use the Ratings service for this experiment and compare \coolName{} variations against HPA-S, AutoBurst and Libra.

\cref{fig:spike-p99} shows the tail (P99) latency during the surge (1600–2200s).
During the surge, AutoBurst's tail latencies remain at the 1s timeout for several minutes,
while HPA-S and Libra oscillate between recovery and repeated timeouts. Such sustained periods of high tail latency degrade user experience.
In contrast, both variants of \coolName{} keep P99 latency stable and within the SLO throughout the surge,
showing no visible impact from the load spike.

We summarize average CPU utilization, SLO violations, and cost in \cref{tab:spike-summary}. Both \coolName{} and \coolName{}-C achieve desirable high CPU utilization of 92\% and 87\% respectively, significantly outperforming HPA-S (58\%), AutoBurst (71\%), and Libra (74\%). Furthermore, the \coolName{} setups nearly eliminate SLO violations (limiting them to just 0.02\%-0.08\%) while simultaneously maintaining the lowest total cost among all compared systems.

\begin{table}[t!]
  \centering
  \begin{tabular}{lrr}
    \toprule
    Offload level & \multicolumn{1}{c}{CPU (\%)} & \multicolumn{1}{c}{Mem. (MB)} \\
    \midrule
     No-offload / Forward-only & 8.1  & 7.3 \\
    10\% offload            & 9.6  & 8.4 \\
    50\% offload            & 13.6 & 9.8 \\
    \bottomrule
  \end{tabular}
  \caption{\slb~ sidecar resource use \emph{per application instance} (averages across instances over the run).}
  \label{tab:overhead-slb}
\end{table}

\subsection{Overhead Analysis}

\label{sec:evaluation:overhead}

\beginbsec{\slb~ Overhead} To evaluate the overhead introduced by the offloading logic in the \slb, we measure the resource usage of the sidecar container colocated with each application container by driving a constant load that keeps ten instances of the Ratings service at $\sim$80\% utilization. \cref{tab:overhead-slb} reports both the average CPU usage (as a percentage of one vCPU) and memory usage of the sidecar across three setups: a baseline “forward-only” proxy that simply forwards traffic to the application container without any offloading logic, and two offloading configurations.
The first is 10\% offload, which serves as a baseline for typical operation,
as \coolName{} offloads at most 5\% of requests in practice (see \cref{sec:evaluation:util-slo}). The second is 50\% offload, capturing worst-case instantaneous spikes under the trace in \cref{fig:load}.

Relative to the baseline, the \slb~ drives CPU usage of the sidecar container from 8.1\% to 9.6\% at 10\% offload (+1.5\%) and from 8.1\% to 13.6\% at 50\% offload (+5.5\%). Memory overhead is +1.1~MB at 10\% offload and +2.5~MB at 50\% offload. We did not observe any measurable increase in request
latency across these setups, indicating that the offloading logic introduces
no noticeable delay beyond the application’s normal processing time. These results show that the \slb’s overhead scales with offloaded volume but remains very small in absolute terms.

\beginbsec{SRO Overhead}
The centralized SRO executes every two minutes, with each iteration completing within a few hundred milliseconds and consuming less than 1\% of a vCPU with negligible memory footprint. This overhead is comparable to other cluster-level autoscaling components such as the k8s HPA controller. Because it runs outside the critical request path, the SRO’s runtime cost is operationally invisible. \Cref{app:sro-scalability} additionally evaluates optimizer runtime for cluster sizes up to 100,000 nodes.

\section{Related work}
\label{sec:related}

\beginbsec{Hybrid Frameworks with Serverless}
A number of works have sought to combine VMs with FaaS. Some target data query tasks~\cite{pixel,splitserve,cackle,sponge}, while others serve cloud workloads~\cite{feat,spock,beehive,mark}. All of these use serverless functions only to process unexpected large load spikes, or to bridge throughput gaps during new VM spin-up. As these systems treat serverless functions only as a fail-safe, they are unable to (1) accommodate fine-grained load variations, and (2) seize the cost-saving opportunities through the use of serverless functions to avoid over-provisioning VMs.

Libra~\cite{libra} goes beyond the aforementioned works in that it continuously offloads to serverless. However, as our work shows, Libra is unable to accommodate fine-grained load variations because its policies conflate provisioning for cost and SLO, ultimately falling short on both fronts. UnFaaSener~\cite{sadeghian2023unfaasener} is a conceptual opposite of \coolName{}. It offloads some serverless functions to VMs when the VM utilization is predicted to be low. Unlike \coolName{}, UnFaaSener is not concerned with SLO and tries to lower cost opportunistically rather than systematically.

\beginbsec{Hybrid Frameworks with VMs}
Numerous prior works seek to optimize cost and performance by combining standard VMs with alternative VM types, such as burstable instances \cite{burscale, dantas2021bias,autoburst_socc} or spot instances~\cite{wu2024cant, mao2025skyserve, miao2024spotserve, xu2023snape, sharma2017portfolio}. Given a highly volatile load, burstable VMs experience frequent use and hence are unable to accumulate credits. To accumulate credits, burstable instances must be under-utilized, which increases cost (\cref{sec:evaluation}).
Spot VMs can be revoked at any time. While cheaper than regular VMs, they share the same limitations as regular VMs when it comes to achieving high utilization under SLO (\cref{sec:sota:autoscale}). Despite the limitations of both spot and burstable VM types, Spandana can leverage them to further lower deployment cost, provide resilience if a VM goes down, and guarantee SLO under fluctuating load.

\beginbsec{Resource allocation for online services}
Prior research has explored resource management for cloud services \cite{sinan, firm, GrandSLAm, pema,ursa,Autothrottle, apollo,jockey,drf,Firmament,Quincy,Morpheus,3Sigma} to balance performance and efficiency. These works focus on efficient distribution of resources (typically, VMs) among a set of services. Spandana operates on top of such cluster-scale resource allocators by ensuring that individual services can meet SLO in the face of fluctuating load.

\section{Conclusion}
\label{sec:conclusion}

To address the sub-second load volatility and the resulting trade-off between SLO compliance and cost efficiency, this work introduced Spandana, a two-level architecture that decouples SLO enforcement from cost optimization. At the VM instance level, a lightweight controller steers requests to the VM if it can meet the SLO target; otherwise, the request is redirected to FaaS. At the global level, Spandana's resource allocator optimizes the entire deployment for cost taking into account VMs, serverless and the shape of the load. Spandana provides strict SLO adherence, high CPU utilization on the VMs and lower cost than SOTA schemes. {\let\thefootnote\relax\footnotetext{Code and artifacts are available at \url{https://github.com/ease-lab/Spandana}}}




\begin{acks}

We thank reviewers and EASE Lab members at the University of Edinburgh for their valuable feedback.
This research was generously supported by the University of Edinburgh, and EASE Lab's industry partners and sponsors, including Huawei, Intel, Arm and Cisco. Marios Kogias is supported by an ERC Starting Grant (ID: 101220079)
\end{acks}
\newpage
\bibliographystyle{ACM-Reference-Format}
\bibliography{bib}

\clearpage
\appendix
\graphicspath{{figures/appendix/}}
\section{Appendix}
\label{app:additional-studies}

This appendix complements the main evaluation with studies of \coolName{}'s resilience, sensitivity, and scalability. We first examine its response to VM failures and evaluate its performance on the more bursty Alibaba trace. We then assess sensitivity to the SRO's load history window and the SLO Director's queue capacity. Finally, we evaluate the SRO's scalability across larger cluster sizes.

\subsection{Handling Node Failures}

\begin{figure}[!t]
    \centering
    \begin{subfigure}[t]{\columnwidth}
        \centering
        \includegraphics[width=\textwidth]{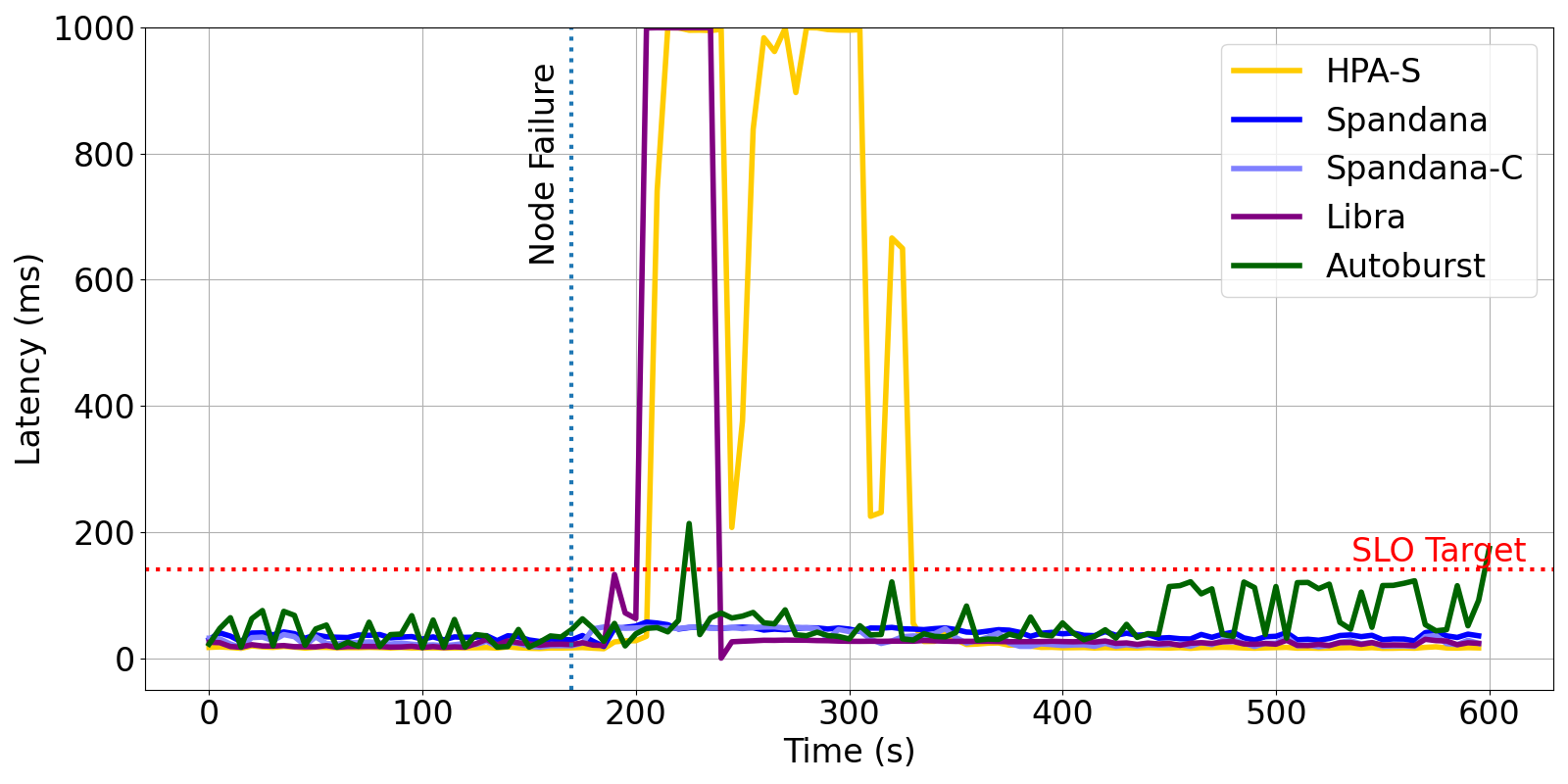}
        \caption{P50 latency}
        \label{fig:node-failure-p50}
    \end{subfigure}

    \vspace{2pt}
    \begin{subfigure}[t]{\columnwidth}
        \centering
        \includegraphics[width=\textwidth]{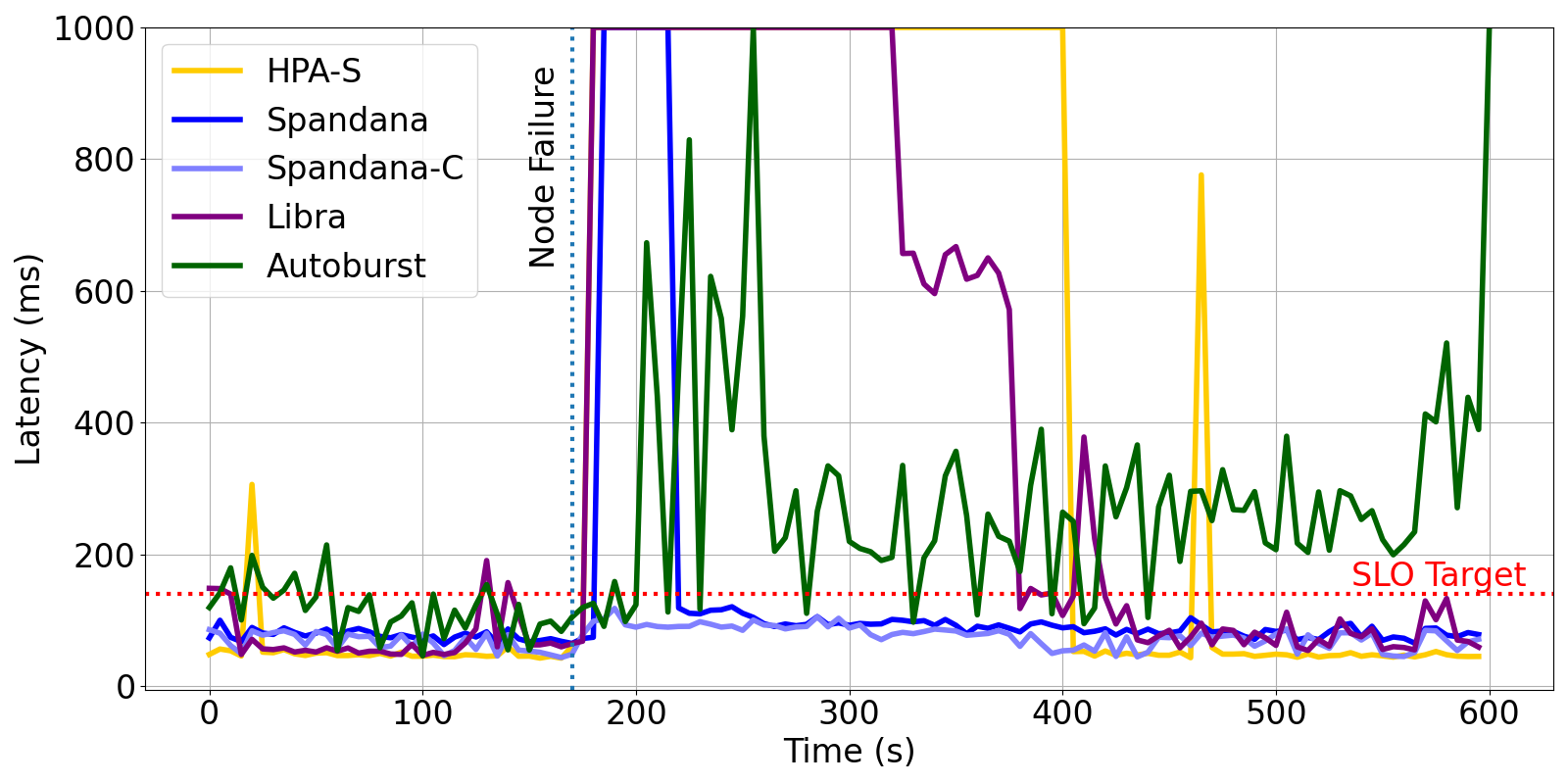}
        \caption{P99 latency}
        \label{fig:node-failure-p99}
    \end{subfigure}
    \caption{Latency response of the \textit{Ratings} service when 50\% of VM instances are manually terminated during a 10-minute Twitter trace.}
    \label{fig:node-failure-latency}
\end{figure}

To evaluate \coolName{} under infrastructure failures, we run the \textit{Ratings} service on a 10-minute Twitter trace and manually terminate 50\% of the VM instances approximately three minutes into the experiment. \Cref{fig:node-failure-latency} shows the resulting median (P50) and tail (P99) latency.

\Cref{fig:node-failure-p50} shows the median latency before and after the node failure. Both \coolName{} variants maintain stable P50 latency after half of the VM instances are terminated. In contrast, HPA-S and Libra experience large latency spikes, while AutoBurst shows smaller spikes but greater variation in median latency than either \coolName{} variant.

\Cref{fig:node-failure-p99} shows the P99 latency. HPA-S and Libra remain above the SLO target for several minutes after the failure. AutoBurst frequently exceeds the target for the remainder of the experiment. \coolName{}'s P99 latency rises briefly because the AWS load balancer takes time to mark terminated nodes as unhealthy and stop routing requests to them. \coolName{}-C uses more frequent health checks (every 2 seconds), allowing it to detect failed nodes sooner and avoid this spike. These results show that both \coolName{} variants maintain stable median latency when nodes fail, while faster failure detection reduces the impact on P99 latency.

\subsection{Evaluation on the Alibaba Trace}

To assess \coolName{} beyond the Twitter workload, we evaluate it on an unscaled Alibaba trace~\cite{madu}. We first characterize the trace's burstiness relative to Twitter, then compare \coolName{} and the baselines in terms of SLO compliance, cost, and CPU utilization on the Ratings service.

\begin{figure}[!t]
\centering
\includegraphics[width=\columnwidth]{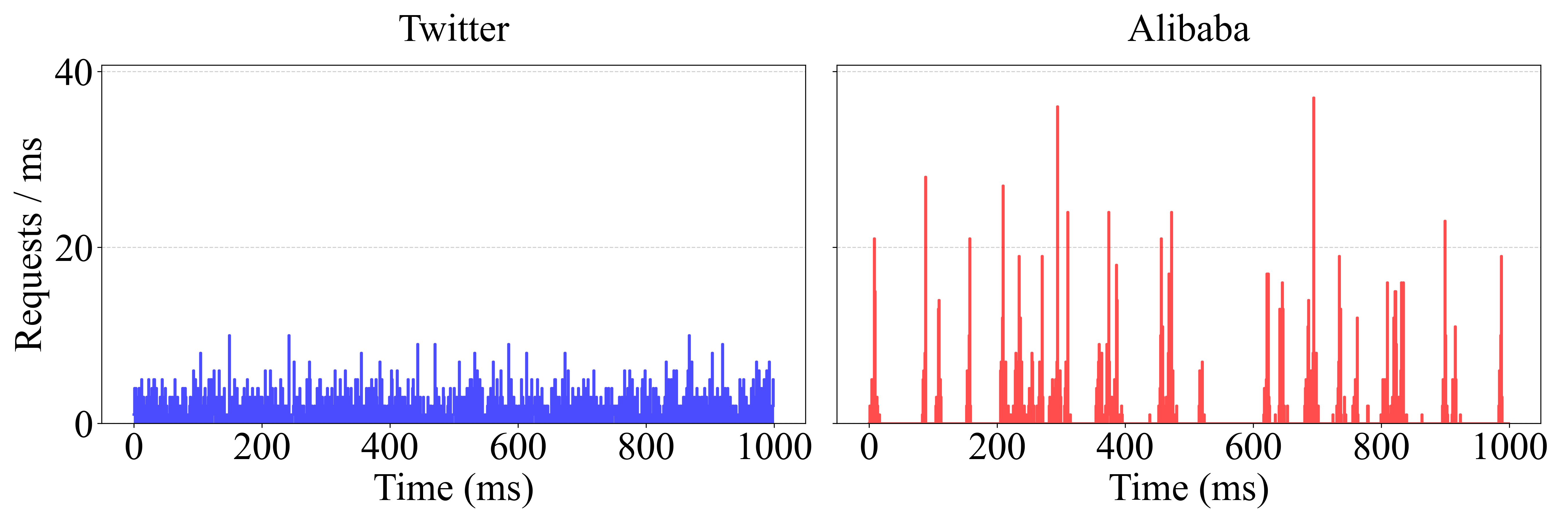}
\caption{Request arrival distributions within a 1-second window.}
\label{fig:trace_comparison}
\end{figure}

\textbf{Trace burstiness:} As shown in \Cref{tab:workload}, the Alibaba trace exhibits substantially higher burstiness than the Twitter baseline, measured by both Peak-to-Mean Ratio (PMR) and Coefficient of Variation (CoV). \Cref{fig:trace_comparison} further visualizes the extreme volatility present within a representative 1-second window.

\begin{table}[!t]
\centering
\small
\setlength{\tabcolsep}{2pt}
\begin{tabular}{@{}lccc@{}}
\toprule
\textbf{Trace} & \textbf{Mean (Req/ms)} & \textbf{CoV} & \textbf{Peak-to-Mean (PMR)} \\ \midrule
Twitter & 1.51 & 0.83 & 7.96 \\
Alibaba & 1.42 & 2.62 & 107.37 \\ \bottomrule
\end{tabular}
\caption{Workload Characterization (1ms Granularity)}
\label{tab:workload}
\end{table}

\begin{table}[!t]
\centering
\small
\begin{tabular}{@{}lccc@{}}
\toprule
\textbf{System} & \textbf{Cost (\textcent)} & \textbf{CPU Util. (\%)} & \textbf{SLO Viol. (\%)}\\
\midrule
Spandana & 67.0 & 71.1 & 0.6 \\
Spandana-C & 64.3 & 72.7 & 0.23 \\
HPA-S & 69.2 & 59.5 & \textcolor{red}{\textbf{1.66}} \\
HPA-O & 83.2 & 42.0 & 0.42 \\
AutoBurst & 53.9 & 76.3 & \textcolor{red}{\textbf{13.72}} \\
Libra & 64.2 & 75.0 & \textcolor{red}{\textbf{12.01}} \\
\bottomrule
\end{tabular}
\caption{Alibaba trace results for Ratings service. \textcolor{red}{\textbf{Red}} indicates violation of the 1\% SLO target.}
\label{tab:alibaba_eval}
\end{table}

\textbf{Efficiency and SLO compliance:} 
The results in \Cref{tab:alibaba_eval} show that both \coolName{} variants meet the 1\% SLO target while achieving low cost. In contrast, all practical SOTA baselines (AutoBurst, Libra, and HPA-S) fail to meet the SLO target due to the highly volatile nature of the trace. HPA-O does meet the SLO target, but is impractical because it uses oracle knowledge of the full trace to provision resources.
The CPU column further highlights \coolName{}'s resource efficiency, with significantly higher utilization than HPA-O.


\subsection{Sensitivity to Load History Window}
\label{app:sro-history-window}
At each optimization step, the SRO uses recent request-rate measurements to estimate the combined VM and serverless cost of candidate provisioning configurations. The \textit{load history window} specifies how far back these measurements extend. A shorter window emphasizes recent traffic, while a longer window captures more historical variation but may include load patterns that are no longer representative. We evaluate window durations of 2 minutes (the default), 5 minutes, and 10 minutes. As shown in \Cref{table:windowsize}, these settings have a negligible impact on total operational cost.

\begin{table}[!t]
\centering
\small
\begin{tabularx}{\columnwidth}{@{}>{\raggedright\arraybackslash}Xccc@{}}
\toprule
 & \multicolumn{3}{c}{\textbf{History Window Size}} \\ \cmidrule(l){2-4} 
\textbf{Metric} & \textbf{2 min (Baseline)} & \textbf{5 min} & \textbf{10 min} \\ \midrule
Serverless Cost (\textcent) & 8.4  & 7.8  & 7.4  \\
VM Cost (\textcent)         & 45.6 & 46.5 & 46.6 \\
\textbf{Total Cost (\textcent)} & \textbf{54.1} & \textbf{54.3} & \textbf{54.0} \\ \bottomrule
\end{tabularx}
\caption{Impact of SRO load history window on operational costs}
\label{table:windowsize}
\end{table}

\subsection{Sensitivity to Queue Length}
\label{app:slo-director-queue-length}
The SLO Director's queue capacity ($Q$) determines the maximum number of requests that can wait locally before newly arriving requests are offloaded to serverless. Larger queues can absorb more bursts locally, reducing offloading costs, but risk longer queueing delays. We evaluate this trade-off on the Ratings service. As shown in \Cref{table:queue-length}, increasing $Q$ reduces cost and offloading, but also increases the likelihood of SLO violations. The default setting of $Q=9$ keeps violations well below the 1\% target while sharply reducing serverless offloading relative to $Q=3$.

\begin{table}[!t]
\centering
\small
\begin{tabularx}{\columnwidth}{@{}>{\raggedright\arraybackslash}Xcccc@{}}
\toprule
\textbf{Metric} & \textbf{$Q=3$} & \textbf{$Q=9$} & \textbf{$Q=15$} & \textbf{$Q=21$} \\ \midrule
SLO Violations (\%)        & 0.04 & 0.05 & 0.82 & \textcolor{red}{4.1} \\
Cost (\textcent)           & 13.9 & 8.9  & 8.6  & 8.3  \\
Requests offloaded to serverless (\%)  & 20.2 & 4.8  & 2.9  & 1.6  \\ \bottomrule
\end{tabularx}
\caption{Sensitivity analysis across different queue lengths ($Q$)}
\label{table:queue-length}
\end{table}

\subsection{SRO Scalability}
\label{app:sro-scalability}
At each optimization step, the SRO evaluates candidate VM counts against the recent load profile to identify the configuration with the lowest estimated deployment cost. Evaluating more candidate VM counts increases the optimizer's computation, while longer load history windows increase the number of data points considered for each candidate. To assess whether this computation remains practical at scale, we measure optimizer execution time on 1 vCPU for cluster sizes of 10K, 50K, and 100K nodes and history windows of 2, 5, and 10 minutes. As shown in \Cref{table:scalability}, execution time remains below 100 ms across all configurations, reaching 77.4 ms at 100K nodes with a 10-minute window. This is small relative to the two-minute optimization interval.

\begin{table}[!t]
\centering
\small
\begin{tabular*}{\columnwidth}{@{\extracolsep{\fill}}lccc@{}}
\toprule
\multirow{2}{*}{\textbf{Node Count}} & \multicolumn{3}{c}{\textbf{Load History Window}} \\ \cmidrule(l){2-4} 
 & \textbf{2 min} & \textbf{5 min} & \textbf{10 min} \\ \midrule
\textbf{10K}  & 1.7 ms  & 3.9 ms  & 7.9 ms   \\
\textbf{50K}  & 8.1 ms  & 19.7 ms & 38.7 ms  \\
\textbf{100K} & 16.6 ms & 39.4 ms & 77.4 ms  \\ \bottomrule
\end{tabular*}
\caption{SRO execution time across cluster sizes and load history windows}
\label{table:scalability}
\end{table}

\end{document}